\documentclass[fleqn,usenatbib]{mnras}

\usepackage{newtxtext,newtxmath}
\usepackage[T1]{fontenc}
\usepackage{ae,aecompl}
\usepackage{algorithm}
\usepackage{algorithmic}
\usepackage{booktabs}
\usepackage{listings}

\usepackage{graphicx}	% Including figure files
\usepackage{amsmath}	% Advanced maths commands
\usepackage{amssymb}	% Extra maths symbols

\title[Fast W-Projection]{Fast W-Projection for Wide-field Imaging}

\author[Lucas, Skipper \& Scaife]{
Lu{\'i}s~F.~R.~Lucas,$^{1,2,3}$\thanks{E-mail: luisfrlucas@gmail.com},
Chris~J.~Skipper$^{1}$
and Anna~M.~M.~Scaife$^{1}$
\\
$^{1}$Jodrell Bank Centre for Astrophysics, Alan Turing Building, University of Manchester, Manchester M13 9PL\\
$^{2}$Critical Software S.A., Coimbra, Portugal\\
$^{3}$Polytechnic Institute of Leiria, Leiria, Portugal
}

\date{Accepted XXX. Received YYY; in original form ZZZ}

\pubyear{2019}

\begin{document}
\label{firstpage}
\pagerange{\pageref{firstpage}--\pageref{lastpage}}
\maketitle

% Abstract of the paper
\begin{abstract}
Wide-field imaging has become a major challenge for  modern radio astronomy, which uses high sensitivity acquisition systems that deal with huge amounts of data. 
In this paper we investigate a fast wide-field imaging solution based on the $w$-projection algorithm, which is intended for modern astronomy systems. The core idea of the proposed method is to reduce the computational complexity of the convolution kernel generation step, specifically by replacing the standard two-dimensional FFT by the one-dimensional Hankel transform.
Experimental results show that the optimised $w$-projection proposed here produces equivalent dirty image results in a circular image region, at a significantly lower computational cost than standard $w$-projection. One of the main advantages of the proposed solution is its slow scaling with the number of $w$-planes, thus enabling more accurate output results at a lower computational cost.
\end{abstract}

% Select between one and six entries from the list of approved keywords.
% Don't make up new ones.

%%%%%%%%%%%%%%%%%%%%%%%%%%%%%%%%%%%%%%%%%%%%%%%%%%

%%%%%%%%%%%%%%%%% BODY OF PAPER %%%%%%%%%%%%%%%%%%

\section{Introduction}

A major obstacle to wide-field imaging in radio interferometry is the non-coplanarity of interferometer baselines. While this effect was not a substantial limitation for early interferometers, which had limited sensitivity and small fields-of-view (FOVs), as the sensitivity of radio telescopes improved over the years the need to implement a full field of view correction became essential.

The main problem caused by the effect of non-coplanar baselines is that it prevents the use of a simple two-dimensional approximation to the transform relationship between the sky brightness distribution, $I$, and the measured visibility data, $V$, \citep{Gridding_1999ASPC} given by:
\begin{equation}
V(u,v) = \int I(l,m) e^{-j2\pi (ul+vm)} {\rm d}l {\rm d}m.
\label{eq:interferom_smallfov}
\end{equation}

In this equation, the sky brightness distribution at a given frequency is represented as a function of the direction cosines, $(l,m)$. The visibility data corresponds to the output of the correlation of a pair of antennas, given as a function of baseline coordinates $u,v,w$ expressed in units of wavelength. 

For larger fields of view, the $w$-term becomes important resulting in a more complex relationship than the 2D Fourier transform:
\begin{equation}
V(u,v,w) = \int \frac{I(l,m)}{\sqrt{1-l^2-m^2}} e^{-j2\pi [ul+vm+w(\sqrt{1-l^2-m^2}-1)]} \operatorname{d}\!l \operatorname{d}\!m.
\label{eq:interferom_general}
\end{equation}
As a consequence, the sky brightness cannot be estimated using a simple Fourier inversion of the measured visibility records. In practice, it can be shown that the $w$-term effect becomes more significant when the field of view is comparable to or greater than the square root of the resolution (both measured in radians) or when the Fresnel number ($N_F$) is less than unity, with:
\begin{equation}
 N_F=\frac{D^2}{B \lambda} ,
\end{equation}
where $B$ is the maximum baseline length, $D$ is the antenna diameter, and $\lambda$ is the observing wavelength. The non-coplanar baseline effect will thus occur for small apertures, long baselines or long wavelengths.

The effect of this non-coplanarity is to attenuate and smear out sources away from the phase centre of the image. A number of different approaches to correcting for this effect have been described in the literature and implemented in radio imaging software. The most accurate of these corrections are highly computationally expensive and, in consequence, significant effort has also been given to deriving more computationally efficient approaches that approximate a full $w$-correction. Here we present a new optimisation of the $w$-projection method \citep{Cornwell_WP2008} using the Hankel Transform for faster performance. Hankel Transforms have previously been proposed to reduce the complexity of Fourier analysis for radially symmetric functions in astronomy before \citep[e.g.][]{birkinshaw1994}, but have not been implemented as part of interferometric image reconstruction algorithms.

The paper is structured as follows, in \S~\ref{sec:approaches} we give a brief overview of existing approaches to $w$-correction; in \S~\ref{sec:base} we describe our reference implementation and in \S~\ref{sec:hankel} we describe our optimized $w$-projection algorithm; in \S~\ref{sec:simulateddata} we describe the simulation data; in \S~\ref{sec:demonstration} we present a comparison between these algorithms; in \S~\ref{sec:comparisonCasapy} we compare our imager performance with CASApy software and in \S~\ref{sec:conclusions} we draw our conclusions.

\section{Overview of W-correction approaches}
\label{sec:approaches}

Various approaches have been proposed in the literature for correcting the effect of the $w$-term. Here we present a brief overview of the currently available algorithms.

\subsection{3D transforms}

The Fourier inversion of Equation \ref{eq:interferom_general} can be written as a three-dimensional Fourier inversion, assuming a 3D space with axes $(l,m,n)$:
\begin{equation}
V(u,v,w) = \int \frac{I(l,m)\delta(n-\sqrt{1-l^2-m^2})}{n} e^{-j2\pi [ul+vm+wn]} \operatorname{d}\!l \operatorname{d}\!m \operatorname{d}\!n.
\label{eq:interferom_3dfourier}
\end{equation}
This equation may be solved using a 3D~FFT or, alternatively, using 2D~FFTs in $(l,m)$ and a DFT in $n$; however, a major drawback of this approach is that a large proportion of the three-dimensional space will be empty, wasting a significant amount of memory.

\subsection{Faceting}

Any wide-field image can be divided into a number of smaller images, usually referred to as facets. For sufficiently small images, the $w$-term can be considered to be near zero, and so facets can be imaged individually using the small-field approximation given by Equation~\ref{eq:interferom_smallfov}.

Each facet is imaged in two dimensions onto a plane that is tangent to the unit sphere at a different point corresponding to the phase center of the individual facet. 
For each facet, both the visibility phases and the $(u,v,w)$ coordinates are then adjusted to a common phase center. This approach is referred to as image-domain faceting, with the small facet images being deconvolved separately before being stitched together~\citep{Cornwell_Faceting1992}.
An alternative is to grid the visibility data multiple times onto the same $uv$-grid, each time for a different phase center, resulting in a single dirty image that is deconvolved entirely using a single PSF~\citep{UVW_facet_Sault1999}. This approach tends to be fast as it avoids dealing with a large number of facet images.

\subsection{Warped snapshots}

For short (snapshot) observations, an array is instantaneously co-planar, which means that the $w$-term can be related to $(u,v)$ by a simple linear relationship \citep{Bracewell_1984}:
\begin{equation}
 w=au+bv.
\label{eq:wterm_plane}
\end{equation}
As a consequence, the $w$-term can be eliminated from Equation~\ref{eq:interferom_general} at the cost of a distortion in the image coordinate system. This coordinate distortion can then be corrected in the image plane for each snapshot using interpolation techniques.

Thus, the relationship between the sky brightness distribution and the visibility may be rewritten as in Equation \ref{eq:interferom_smallfov} by introducing distorted coordinates $(l',m')$ where:
\begin{equation}
\begin{aligned}
 l' &= l + \tan(Z)\sin(\mathcal{X}) (\sqrt{1-l^2-m^2}-1)\\
 m' &= m - \tan(Z)\cos(\mathcal{X}) (\sqrt{1-l^2-m^2}-1)
\end{aligned}
\end{equation}
where $Z$ is the Zenith angle, and $\mathcal{X}$ is the parallactic angle at the time of observation. 

A major issue for this approach is the large computational complexity due to the high accuracy required for the image re-projection. However, when used in conjunction with $w$-projection, an optimum trade-off between the computational resources needed for each method can be achieved, see \S~\ref{ssec:wsnap}.

\subsection{W-stacking}

The $w$-stacking approach grids visibility records onto different $w$-layers and performs the $w$-correction {\it after} the inverse Fourier transform. The $w$-stacking method can be easily derived from Equation~\ref{eq:interferom_general}, which can be rewritten as:
\begin{eqnarray}
\nonumber \frac{I'(l,m) (w_{max} - w_{min})}{\sqrt{1-l^2-m^2}} &=& \int_{w_{min}}^{w_{max}} e^{j2\pi w (\sqrt{1-l^2-m^2}-1)} \\
\nonumber &\times& \int\int V(u,v,w)e^{j2\pi (ul+vm)} \operatorname{d}\!u \operatorname{d}\!v \operatorname{d}\!w. \\
&&
\label{eq:wstacking}
\end{eqnarray}
This equation shows that the sky image can be obtained using the following steps:
\begin{enumerate}
 \item Grid visibility data with equal $w$-terms;
 \item Calculate the inverse FFT;
 \item Multiply by the phase shift $e^{j2\pi w (\sqrt{1-l^2-m^2}-1)}$;
 \item Repeat the previous steps for all $w$-terms and add the results together;
 \item Apply the final scaling $\frac{(w_{max} - w_{min})}{\sqrt{1-l^2-m^2}}$
\end{enumerate}
More details about the $w$-stacking algorithm are given in \cite{Offringa_WStacking2014}.

\subsection{W-projection}
\label{ssec:wprojection}

While $w$-stacking expresses the $w$-term as a multiplicative effect in the image domain, $w$-projection exploits the fact that the $w$-term is a convolution in Fourier space. 
In fact, $w$-projection re-projects visibility data from any position in $(u,v,w)$ space to the $w=0$ plane using a convolution with a known kernel. This result can be derived by rewriting Equation~\ref{eq:interferom_general} as:
\begin{equation}
V(u,v,w) = \int \frac{I(l,m)}{\sqrt{1-l^2-m^2}} G(l,m,w) e^{-j2\pi [ul+vm]} \operatorname{d}\!l \operatorname{d}\!m,
\label{eq:wproj_der}
\end{equation}
with:
\begin{equation}
G(l,m,w) = e^{-j2\pi [w(\sqrt{1-l^2-m^2}-1)]}.
\label{eq:wproj_G}
\end{equation}
Using the convolution theorem, Equation~\ref{eq:wproj_der} can then be written as:
\begin{equation}
V(u,v,w) = \tilde{G}(u,v,w) \ast V(u,v,w=0) ,
\label{eq:wproj_convthe}
\end{equation}
where $\tilde{G}(u,v,w)$ is the Fourier transform of $G(l,m,w)$.

This equation explicitly shows the convolution relationship between any position in the $(u,v,w)$ space and the $w=0$ plane, and illustrates that interferometers with the same $(u, v)$ but different $w$-terms provide substantially different information about the sky brightness. An interesting interpretation of $w$-projection is given by Fresnel diffraction theory presented in \cite{Cornwell_WP2008}.

\subsection{W-snapshots}
\label{ssec:wsnap}
 
The $w$-snapshots approach combines the $w$-projection and warped snapshots methods so that an optimum trade-off in computational resources can be obtained \citep{Cornwell_SKA_2012}.
This method expresses the $w$-term as a linear plane, as given in Equation~\ref{eq:wterm_plane} for the warped snapshots approach, plus a deviation $\Delta w$, resulting in:
\begin{equation}
 w=au+bv+\Delta w.
 \label{eq:wterm_planedev}
\end{equation}
The best choice of plane in $u,v,w$ space is chosen using a least-squares fit, and $w$-projection is used to correct the residual $\Delta w$ by projecting all the visibility records onto the plane. Snapshot imaging is then performed. Plane fitting is repeated when the deviation from the previous best-fit plane exceeds a specified tolerance.

\section{Base Algorithm}
\label{sec:base}

Interferometric imaging revolves around the use of the Fast Fourier Transform (FFT) in order to minimize computational complexity. It is for this reason that the measured visibility data from radio interferometers, which are natively sampled in time and frequency, are gridded onto a regularly spaced array of spatial frequencies, $(u,v)$, before being transformed into the image domain.

The gridding operation requires the use of an anti-aliasing (AA) gridding convolution function (GCF) in order to avoid artefacts appearing in the output image due to the periodic nature of the input grid. As a consequence, the output dirty image obtained after FFT will also require a correction to be made in order to compensate for this operation. This correction is simply the division of the output dirty image by the FFT of the GCF.

The form of the AA-kernel can be any function with non-zero compact support within the required FOV of the output image; the 2D Gaussian function can be used for this purpose, but the most commonly used AA-kernel is the prolate spheroidal wavefunction (PSWF). Typically, the AA-kernel is found to be a separable function, i.e. it can be written as the outer product of two vectors, which reduces both the computational and memory requirements of its use. 

In practice, the implementation of the $w$-projection algorithm described in \S~\ref{ssec:wprojection} consists in modifying the GCF. If the $w$-term is significant, the GCF is modified to include its effect. This consists in multiplying $G(l,m,w)$ (see Eq.~\ref{eq:wproj_G}) with the Fourier transform of the AA function and then transforming the result back to the Fourier space to create a modified GCF. A description of the standard $w$-projection algorithm is given in Algorithm~\ref{alg:wprojection}.

\begin{algorithm}
\begin{algorithmic}[1]
 \STATE Sort input visibility records by increasing $w$-term. \label{step:sort}
 \STATE Divide visibility records into M equally-sized $w$-planes based on sorted $w$-terms.
 \STATE Compute the average (or median) $w$-term for each $w$-plane. \label{step:averagew}
 \STATE Compute image-domain AA-kernel (IAA-kernel) in workspace with image size. \label{step:aakernel}
 \FOR{each $w$-plane \label{step:for}}
    \STATE Compute $w$-kernel, $G(l,m,w)$, for current $w$-term (use workspace with image size). \label{step:wkernel}
    \STATE Compute image-domain convolution kernel (IConv-kernel) by multiplying $w$- and IAA-kernels. \label{step:multiply}
    \IF{oversampling $> 1$}
    \STATE Pad IConv-kernel with zeros (increase size by a factor equal to oversampling ratio). \label{step:padding}
    \ENDIF
    \STATE Determine convolution kernel (Conv-kernel) by computing FFT of IConv-kernel. \label{step:convkernel}
    \STATE Truncate Conv-kernel (typically at 1\%). \label{step:truncate} 
    \IF{oversampling $> 1$ \label{step:genkernelcache}}
    \STATE Extract kernels at oversampled positions and build kernel-cache.
    \ENDIF
    \STATE Grid visibility records assigned to the current $w$-plane using Conv-kernel (or the kernel cache in the oversampling case). \label{step:grid}
 \ENDFOR
\end{algorithmic}
\caption{Detailed algorithm of standard $w$-projection.}
\label{alg:wprojection}
\end{algorithm}

Ideally, a distinct GCF should be computed for each $w$-term. Unfortunately, such an approach would be extremely slow, since the GCF generation from the AA- and the $w$-kernels is computationally expensive. To reduce the kernel generation overhead, the standard $w$-projection algorithm \citep{Cornwell_WP2008} divides the input visibility data into a limited number of $w$-planes which are gridded using the same GCF, reducing thus the total number of convolution kernels required. For each $w$-plane, the GCF is determined using the average or median $w$-term of the $w$-plane. 

Steps \ref{step:sort} to \ref{step:averagew} of Algorithm~\ref{alg:wprojection} correspond to the computation of $w$-planes. Initially, the visibility data are sorted by increasing $w$-term for posterior determination of $N_{w}$ equally-sized $w$-planes. 
Using the sorted data is also advantageous for faster computation of the median or average $w$-term within each $w$-plane. 
Furthermore, since the convolutional gridding step processes each $w$-plane at a time, accessing the data sequentially from the sorted array has performance benefits. The arrays sorted by increasing $w$-term include the visibility records, the $uvw$ data and the corresponding weight values.

When using $w$-projection it can be impractical to generate and store the entire kernel cache of all $w$-planes before the gridding step due to the large number of kernels involved. For example, without $w$-projection, the kernel cache for an oversampling ratio of 8 would contain 64 different kernels. In the case of $w$-projection, assuming the same oversampling ratio and $N_{w} = 100$ $w$-planes, a cache of 6400 kernels would be required, which is two orders of magnitude larger. Additionally, the memory used per kernel would also be significantly larger as the support for the $w$-kernels is typically larger than for the AA-kernel only. Thus, the kernel cache is typically computed for each $w$-plane.

The \textit{for} cycle in step~\ref{step:for} of Algorithm~\ref{alg:wprojection} reflects the kernel cache update per $w$-plane. The iteration over each $w$-plane requires computing the GCF, building the kernel cache (when oversampling is used) and performing the convolutional gridding. The AA-kernel generation (step~\ref{step:aakernel} of Algorithm~\ref{alg:wprojection}) is the only step that can be performed once before the gridder loop.

\subsection{Convolution kernel generation}
\label{ssec:conv_kernel_gen}

As previously described, the $w$-projection GCF is built from the AA-kernel and $w$-kernel. Two main approaches can be used to combine these kernels:
\begin{enumerate}
 \item Multiply image-domain AA- and $w$-kernels and compute the DFT of the result;
 \item Convolve the Fourier-domain AA- and $w$-kernels.
\end{enumerate}
Given that the convolution operation tends to be computationally complex, common $w$-projection algorithms use the first approach, which multiplies the image-domain kernels and transforms the result to the Fourier domain.
However, the additional Fourier transforms required by the first approach can also be computationally expensive.
Typically, the image-domain AA- and $w$-kernels are computed in a workspace area equivalent to the size of the image. The image-domain AA-kernel only needs to be computed once, since it does not change in the gridding procedure.

After multiplying the image-domain AA- and $w$-kernels, an FFT is applied to obtain the convolution kernel in the Fourier domain. $w$-kernel computation, kernel multiplication and FFT are each performed within the gridding loop for each $w$-plane, as represented in Algorithm~\ref{alg:wprojection} by steps~\ref{step:wkernel} to~\ref{step:convkernel}.
To limit the size of the convolution kernel, 1\% kernel truncation is typically applied (see step~\ref{step:truncate} of Algorithm~\ref{alg:wprojection}). 
Kernel truncation consists of limiting the kernel size to an extent where the absolute amplitude of the function becomes less than 1\% of its maximum (at the function centre).

As the convolution kernel is generated through the use of an FFT, which is intrinsically discrete, it is not possible to produce kernels at exact $uvw$ positions. Kernel oversampling is therefore used to improve the accuracy of $w$-projection. Over-sampling is achieved by zero padding the multiplied AA- and $w$-kernels (before the FFT step) by an amount equal to the oversampling ratio, as described in step~\ref{step:padding} of Algorithm~\ref{alg:wprojection}. 

A major issue associated with the kernel oversampling is the increase in computational complexity caused by the larger FFT size. For example, for an oversampling ratio of 8, the FFT size is 8 times larger than the image size, which according to $\mathcal{O}(N^2 \log_2{N^2})$ asymptotic complexity (for $N\times N$ function) has a high impact on the computational complexity. 

It is thus crucial to investigate efficient solutions that minimise the cost of the convolution kernel generation in $w$-projection.

\section{Optimised Algorithm}
\label{sec:hankel}

From a starting point of our base algorithm we now describe the new fast $w$-projection algorithm which forms the focus of this paper.
The main optimisation techniques described here involve the use of absolute $w$-terms, workspace undersampling and the Hankel transform.

\subsection{Absolute $w$-term}
\label{ssec:absolutewterms}

The first optimisation we propose is to compute convolution kernels using the absolute $w$-terms, exploiting the fact that the negative $w$-term produces a convolution kernel that is the conjugate of that with a positive $w$-term. 
Given that the visibility data always contain pairs of positive and negative $w$-terms, this optimisation allows one to reduce the number of computed convolution kernels by a factor of two.
Furthermore, it simplifies data partitioning into $w$-planes, since the data sorting and $w$-plane determination can be performed using the absolute $w$-terms.

\subsection{Workspace undersampling}
\label{ssec:undersampling}

As explained in previous section, the GCF is generated in a workspace with the same size as the image, for an accurate result. 
However, the 1\% GCF truncation usually occurs at a significantly smaller size (in the Fourier domain), since the amplitude of the GCF tends to drop quickly to small values around zero.
Given that the size of the GCF in the Fourier domain is usually significantly smaller than the image size, assuming that the GCF is zero beyond the truncation size, we can reduce the workspace size in the image domain (where image-domain AA- and $w$-kernels are multiplied) to the same the truncated size for a faster and approximately equivalent GCF computation.

To exploit this optimisation, it is thus necessary to know the truncated size of the GCF. Since this varies with $w$-plane and is not known before the Fourier transform is performed, we added an input parameter to the imager which defines the maximum allowed convolution kernel size. The maximum GCF size must thus be defined to be a sufficiently large value, ensuring that it is not less than the largest GCF truncation size. Otherwise, relevant GCF values with non-negligible amplitude may be discarded, affecting the computed GCF accuracy. 

We called this method workspace undersampling, because it reduces the workspace size of the AA- and $w$-kernels by undersampling these kernels. In fact, we are doing exactly the opposite of the oversampling technique which uses zero padding in one domain to get an oversampled function in the other domain. Here, we are undersampling a function in one domain to remove zero padding (all values beyond the truncation size) in the other domain.

By reducing the workspace size, we obtain significant computational performance gains, given that the number of points required to compute the AA- and $w$-kernels is much smaller. Furthermore, the FFT size that generates the GCF is significantly smaller. Due to the fact that our C++ implementation of the algorithm uses pre-optimised FFTW \citep{Frigo_FFTW_website, Frigo_FFTW05} plans for power of two sizes, the algorithm always approximates the undersampled workspace size to the next highest  power of two size.

\subsection{Hankel transform}
\label{ssec:hankeltransform}

Unlike the AA-kernel, the $w$-kernel is not separable, in spite of being a radially symmetric function. The same applies to the product of the AA- and $w$-kernels, which also cannot be written as the outer product of two vectors. This fact prevents the potentially large gain in performance that could be obtained by replacing the 2D-FFT by a single 1D-FFT. 
However, although it is not separable, the product of the AA- and $w$-kernels does possess a radial symmetry. Here we exploit this symmetry for better performance by using the Hankel (or Bessel) transform (HT) \citep{Piessens_HT2000}.

The Hankel transform of order $n$ provides an efficient solution to transforming radially symmetric functions. In particular, the zeroth order HT is equivalent to the two-dimensional Fourier transform of a radially symmetric function. This equivalence can be demonstrated using the Fourier transform definition by introducing polar coordinates and considering the fact that radially symmetric functions are independent of $\theta$. 

The Hankel transform of order $\nu$, $G_\nu(k)$, expresses any given function, $g(r)$, as the weighted sum of an infinite number of Bessel functions of the first kind, $J_\nu(kr)$, such that,
\begin{equation}
    G_\nu(k) = \int_0^\infty g(r)J_\nu(kr)\,r\operatorname{d}\!r.
    \label{eq:ht}
\end{equation}
For a more detailed discussion of Hankel transforms please see \cite{Piessens_HT2000}.

The approximate Hankel Transform of a given discrete signal can be obtained using the discrete HT transform matrix, by multiplying this matrix and the input signal represented as a vector. Ignoring the HT matrix computation (which only needs to be done once), this solution has an asymptotic complexity of $\mathcal{O}(N^2)$ due to the matrix-vector product, which is less than the 2D FFT asymptotic complexity of $\mathcal{O}(N^2 \log_2(N^2))$. Consequently, even a naive implementation of this approach should therefore have superior performance characteristics; however, here we present an even more efficient implementation of the Hankel transform, which exploits the use of the projection-slice theorem and is described as follows.

\subsubsection{Projection-slice theorem}

The projection-slice theorem states that the one-dimensional Fourier transform of the projection of a given function $f(x,y)$ onto a line in the $x-y$ plane at any angle is a slice of $F(u,v)$ (the Fourier transform of $f(x,y)$) along a radial line in the $u-v$ plane at the same angle.
Considering a slice of $F(u,v)$ at $v=0$, or equivalently, $G_\nu(k)$, the Hankel transform in Equation~\ref{eq:ht} can be rewritten according to projection-slice theorem as:
\begin{equation}
 G_\nu(k) = \int_{-\infty}^{\infty} e^{j2\pi u x} p(x) dx\;,
\end{equation}
where $p(x)$ is the projection of $f(x,y)$ given by:
\begin{equation}
 p(x) = \int_{-\infty}^{\infty} f(x,y) dy\;.
\end{equation}

From these equations, we conclude that the HT can be implemented using a 1D FFT and function projection, which corresponds to an asymptotic complexity of $\mathcal{O}(N \log_2(N))$, due to the 1D FFT. Note that the projection step in discrete time does not involve a multiplication, but is obtained from $N^2$ sums. This complexity contrasts with the previously presented complexity of the matrix-vector HT method of $\mathcal{O}(N^2)$.

\subsubsection{2D-kernel interpolation}

A consequence of using the HT is that the convolution kernel is generated in one-dimension (a radial line). 
However, for the $uv$-gridding process a two-dimensional convolution kernel is required. 
To achieve this we generate the 2D-kernel by interpolating the radial kernel line produced by the Hankel transform.

In our algorithm, interpolation can be performed using a linear or a cubic spline. Due to the reduced number of calculations involved, the linear spline performs faster than a cubic spline; however, the accuracy of the output image may be superior with cubic spline interpolation. We implemented both interpolation methods using an efficient solution that first computes all the interpolation coefficients for the available points, so that any given point can be immediately interpolated from the pre-computed coefficients.

\subsection{Optimised $w$-projection algorithm}

A detailed description of the optimised $w$-projection algorithm developed here is shown in Algorithm~\ref{alg:opt_wprojection}. 
One of the main advantages of Hankel Transform optimisation is the fact that kernel oversampling may be performed using significantly less memory. Rather than zero padding a 2D kernel, as is done in the standard $w$-projection algorithm, here zero padding is applied to the one-dimensional projected function. Also, as kernel oversampling directly affects the Fourier transform size, its impact on the computational complexity is significantly inferior in the optimized approach which uses a 1D FFT.

The output image from our optimised $w$-projection differs from the standard approach, producing a circularly shaped dirty image, instead of a square image. The circular dirty image shape is a consequence of the 2D interpolation of the convolution kernel by radial rotation, and corresponds to a circular image-domain AA-kernel. This affects the image correction step which attenuates the regions outside a circular image region, compared to using the square shaped AA-kernel (see \S~\ref{sec:defaultexperiments}).

\begin{algorithm}
\begin{algorithmic}[1]
 \STATE Sort input visibility records by increasing absolute $w$-term.
 \STATE Divide visibility records into M equally-sized $w$-planes based on corresponding sorted absolute $w$-terms.
 \STATE Compute the average (or median) $w$-term for each $w$-plane.
 \STATE Determine the undersampled workspace size.
 \STATE Compute image-domain AA-kernel (IAA-kernel) using workspace size.
 \FOR{each $w$-plane}
    \STATE Compute $w$-kernel for current $w$-term (use workspace size).
    \STATE Compute image-domain convolution kernel (IConv-kernel) by multiplying $w$-kernel and IAA-kernel.
    \STATE Compute 1D projection function of IConv-kernel.
    \IF{oversampling $> 1$}
    \STATE Pad projected function with zeros (increase size by a factor equal to the oversampling ratio).
    \ENDIF
    \STATE Determine 1D convolution kernel (Conv-kernel) by computing 1D-FFT of the zero padded projected function.
    \STATE Truncate 1D Conv-kernel at selected percentage (default is 1\%).
    \STATE Interpolate 2D Conv-kernel from 1D array, using linear or cubic splines.
    \IF{oversampling $> 1$}
    \STATE Extract kernels at oversampled positions and build kernel-cache.
    \ENDIF
    \STATE Grid visibility records assigned to the current $w$-plane using 2D Conv-kernel (or the kernel cache in the oversampling case). 
 \ENDFOR
\end{algorithmic}
\caption{Optimised $w$-projection algorithm using Hankel transform.}
\label{alg:opt_wprojection}
\end{algorithm}

\section{Simulation Settings}
\label{sec:simulateddata}

In order to demonstrate the optimized $w$-projection algorithm, we simulated visibility data affected by  non-coplanar baselines effects.
The telescope configuration used was based on the one described in \cite{Cornwell_WP2008}, which has been used with sixty six sources taken from the Westerbork Northern Sky Survey (WENSS) \citep{WENSS_1997}.
The main details of this configuration are:
\begin{itemize}
 \item 74 MHz VLA-C telescope (a low frequency observation increases the non-coplanar baseline effect);
 \item Pointing centre at J\,12h56m57.2 +47d20m20.8;
 \item Measurement set of 505440 visibility records.
\end{itemize}

For the simulated source distributions, two different scenarios were used:
\begin{description}
 \item[\textbf{WENSS}] 66 sources taken from the Westerbork Northern Sky Survey with a brightness larger than 2\,Jy within 12 degrees of the specified centre (based on \cite{Cornwell_WP2008}). The brightest source has a strength of 47.8\,Jy and has been chosen to lie at the field center.
 \item[\textbf{5DIAG}]  5 sources with a strength of 1\,Jy each, positioned on the image diagonal with equally spaced positions from the field centre to the bottom-right field corner with the furthest source at a radial distance of 12\,degrees.
\end{description}

The standard and optimised $w$-projection approaches were implemented in C++ in the context of the SKA Science Data Processor (SDP) Slow Transients Pipeline (STP) design demonstrator software  \citep{Lucas_STPSet2017}, available online in GitHub \citep{STP_website}. The implementation uses multi-core processing based on the Intel Threading Building Blocks (TBB) \citep{TBB_site}.
For this paper, the STP software has been compiled using double-precision floating-point representation, although it can also be compiled using single-precision format for faster execution.
The experiments were performed using an Intel i7-3720QM CPU machine running at 2.6\,GHz, with 32\,GB RAM, hyper-threading disabled and running the Debian 9.0 operating system.

The following standard $w$-projection settings were used for both the \emph{WENSS} and \emph{5DIAG} simulated data sets:
\begin{lstlisting}
    num_wplanes: 128,
    wplanes_median: false,
    max_wpconv_support: 127,
    hankel_opt: false,
    hankel_proj_slice: false,
    undersampling_opt: 1,
    kernel_trunc_perc: 1.0,
    interp_type: cubic
\end{lstlisting}

The above configuration is used for the standard $w$-projection setting that uses a 2D FFT to generate the convolution kernels, which we refer to as \emph{WP-FFT}. In addition, we define the \emph{WP-Hankel} configuration which differs from \emph{WP-FFT} only by the ``hankel\_opt'' and ``hankel\_proj\_slice'' settings, which are set to  true. 
By default, the number of $w$-planes is set to 128 in both configurations. It is important to note that in the case of STP, this is the number of $w$-planes computed in the positive $w$-domain. In practice, as the negative $w$-planes are computed from the positive $w$-planes, the total number of $w$-planes actually used is equivalent to twice the configuration setting (e.g. STP uses 256 in this case). The second setting indicates that each $w$-plane is computed from the average $w$-values. The maximum convolution kernel support size is set to 127, which corresponds to a full kernel size of $255\times 255$.
Workspace undersampling is enabled for both configurations and the kernel truncation percentage is set to $1\%$. In the case of the \emph{WP-Hankel} configuration the kernel interpolation type is set to cubic.

In practice, the difference between \emph{WP-FFT} and \emph{WP-Hankel} configurations is essentially the use of the Hankel transform optimisation. Both configurations use the absolute $w$-term optimisation (\S~\ref{ssec:absolutewterms}) and workspace undersampling (\S~\ref{ssec:undersampling}). In fact, the use of workspace undersampling is essential, otherwise the volume of memory and computational complexity used would be prohibitive. For instance, using an image size of $16384\times 16384$ and oversampling of $8$ would result in a zero padded workspace size of $131072 \times 131072$ pixels (a 275\,GB array of complex data).

In addition to the $w$-projection settings, some imager settings need to be defined. In these experiments, two sets of imager configurations were used, namely:
\begin{itemize}
 \item IM-S - Image size $2048\times 2048$ and cell size of $60$\,arcsec; 
 \item IM-L - Image size $16384\times 16384$ and cell size of $7.5$\,arcsec.
\end{itemize}

These configurations correspond to a large and small image size setting, using the same field-of-view size, a consequence of the distinct cell sizes. The imager settings also include a default oversampling ratio of 8 and an AA-kernel defined as a Gaussian function.
It is important to refer that the Gaussian convolution kernel is chosen as an alternative to the widely used prolate spheroidal wave function (PSWF) due to the radially symmetric property of the 2D Gaussian, obtained by the cross product of two 1D Gaussian functions.
This is because the 2D PSWF is not exactly radially symmetric when computed from the cross product of 1D functions. Radial symmetry could be enforced using interpolation methods to generate 2D PSWF, but this approach would be more complex and less precise than using the separable representation of the Gaussian kernel.

Note also that STP uses pre-computed FFTW plans for faster FFT execution in the target machine. These plans were pre-generated only for power of two matrix sizes. It would be impractical to pre-compute optimised plans for thousands of possible matrix sizes. Due to this restriction, the workspace size used for oversampled convolution kernel generation is set to a power of two value determined from the maximum support size specified in the $w$-projection configuration.

\section{Comparison to standard $w$-projection}
\label{sec:demonstration}

In this section we evaluate both standard and optimised $w$-projection algorithms implemented in the context of the STP, referred to \emph{WP-FFT} and \emph{WP-Hankel}, respectively. It is important to note that the STP imager performs only the gridding and inverse Fourier steps to produce a dirty image, and thus does not implement the CLEAN operation. 

In these experiments we analyse the influence of each $w$-projection setting on the run time and source peak amplitude, by plotting these metrics as a function of some varying parameter value. 
The source peak amplitude has been detected using the source find algorithm proposed in \cite{Lucas_SourceFind2019} for the STP, using analysis and detection thresholds equal to 20\,$\sigma$.

For the benchmark tests we measure the run time in seconds. Although the absolute run time may vary significantly between different machines and depends on the implementation, it does provide a realistic measure of the algorithm's computational complexity. 
This statement can be justified by the fact that the number of operations alone does not reliably estimate the computational complexity. For instance, cache reuse and locality, and access patterns to main memory are also important aspects of an algorithm which are not reflected in the number of operations. Sometimes, algorithms that perform fewer operations might require more complex memory access patterns, and thus be far slower than alternative implementations requiring more operations. 
Additionally, often the number of operations can be reduced by using cached data. Thus, in order to account for all these aspects in the computational complexity evaluation, we opt to measure the execution time of the algorithms, using a C++ implementation which has been optimised to the best of our knowledge.

For most benchmarks, in addition to the total run time of the imager (gridding + FFT), we provide the time used for the gridding step, as well as the time used for convolution kernel generation, which is part of the gridding step, separately.
To evaluate the accuracy of the output dirty image, we measure the peak amplitude in Janskys for each detected source, comparing it with the expected amplitude. For simplicity, the amplitude results are demonstrated using the 5DIAG data, which contains a limited number of sources distributed at different distances from the field centre with equal expected peak amplitudes of 1.0\,Jy.

\subsection{Results using default settings}
\label{sec:defaultexperiments}

In this subsection, we provide the first analysis of the imager results, when not using $w$-projection (\emph{WP-Off}), using the standard $w$-projection (\emph{WP-FFT}) and using the optimised $w$-projection (\emph{WP-Hankel}) for the default settings presented in \S~\ref{sec:simulateddata}. We used the \emph{IM-L} imager configuration (image size of $16384\times 16384$).

\begin{table*}
\centering
\caption{STP imager run times (in seconds) and source peak amplitude results (in Jansky) for \emph{IM-L} imager setting ($16384\times 16384$) and three $w$-projection configurations: $w$-projection disabled (WP-Off), using standard $w$-projection (WP-FFT) and using optimised $w$-projection (WP-Hankel).}
\label{tab:imager_bench}
\begin{tabular}{|l|c|c|c|c|c|c|}
\hline
\textbf{Config} & \textbf{Run Time {[}s{]}} & \textbf{S1 [Jy]} & \textbf{S2 [Jy]} & \textbf{S3 [Jy]} & \textbf{S4 [Jy]} & \textbf{S5 [Jy]} \\ \hline
WP-Off     & 2.099   & 1.00072 & 0.37259 & Not detected & Not detected & Not detected \\ \hline
WP-FFT     & 11.514  & 1.0007 & 1.00189 & 0.970408 & 0.964346 & 0.928649  \\ \hline
WP-Hankel  & 6.479   & 1.00077 & 1.00176 & 0.970096 & 0.96484 & 0.931308  \\ \hline
\end{tabular}
\end{table*}

Table \ref{tab:imager_bench} presents the run time and source peak amplitude results for the 5DIAG measurement set. We observe that the STP imager takes approximately 2 seconds to generate a $16384 \times 16384$ image on the 4-core machine when $w$-projection is disabled. The use of $w$-projection increases the computational complexity as expected,  but the increase for the optimised \emph{WP-Hankel} is smaller, being about 3 times slower than \emph{WP-Off}, in contrast with the standard approach \emph{WP-FFT}, which is about 5.5 slower.
As we will discuss in the rest of the paper, the potential gain of \emph{WP-Hankel} relative to \emph{WP-FFT} is increasingly superior when a larger number of $w$-planes is used.
Regarding the use of the WENSS measurement set, we observed the same run time results as the ones for 5DIAG. Note that both 5DIAG and WENSS measurement sets have the same number of visibility records, differing just in the source distribution.

The source peak amplitude results for the 5DIAG data are presented in the last 5 columns of Table \ref{tab:imager_bench}. S1 corresponds to the source at the field centre while S5 corresponds to the furthest source. We may observe that both \emph{WP-FFT} and \emph{WP-Hankel} present similar peak amplitude results, close to the expected value of 1\,Jy, and the amplitude error increases with  distance to the field centre. When $w$-projection is disabled, the peak amplitude of the sources strongly decays with the distance to the field centre and the furthest 3 sources are not even detected due to smearing.

\begin{figure*}
    \centering
    \includegraphics[width=0.48\textwidth, clip, trim=0cm 2cm 0cm 2cm]{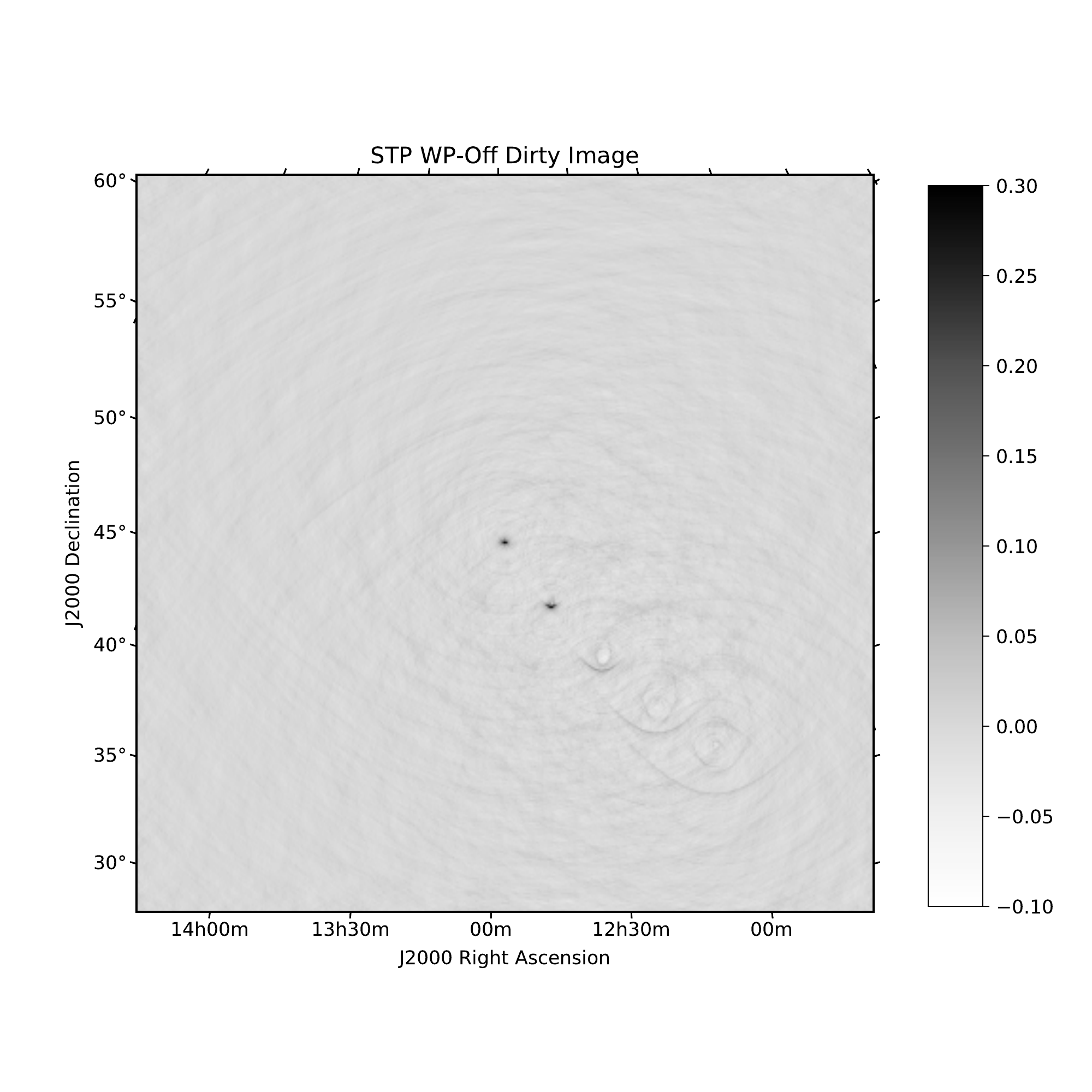}\quad\quad
    \includegraphics[width=0.48\textwidth, clip, trim=0cm 2cm 0cm 2cm]{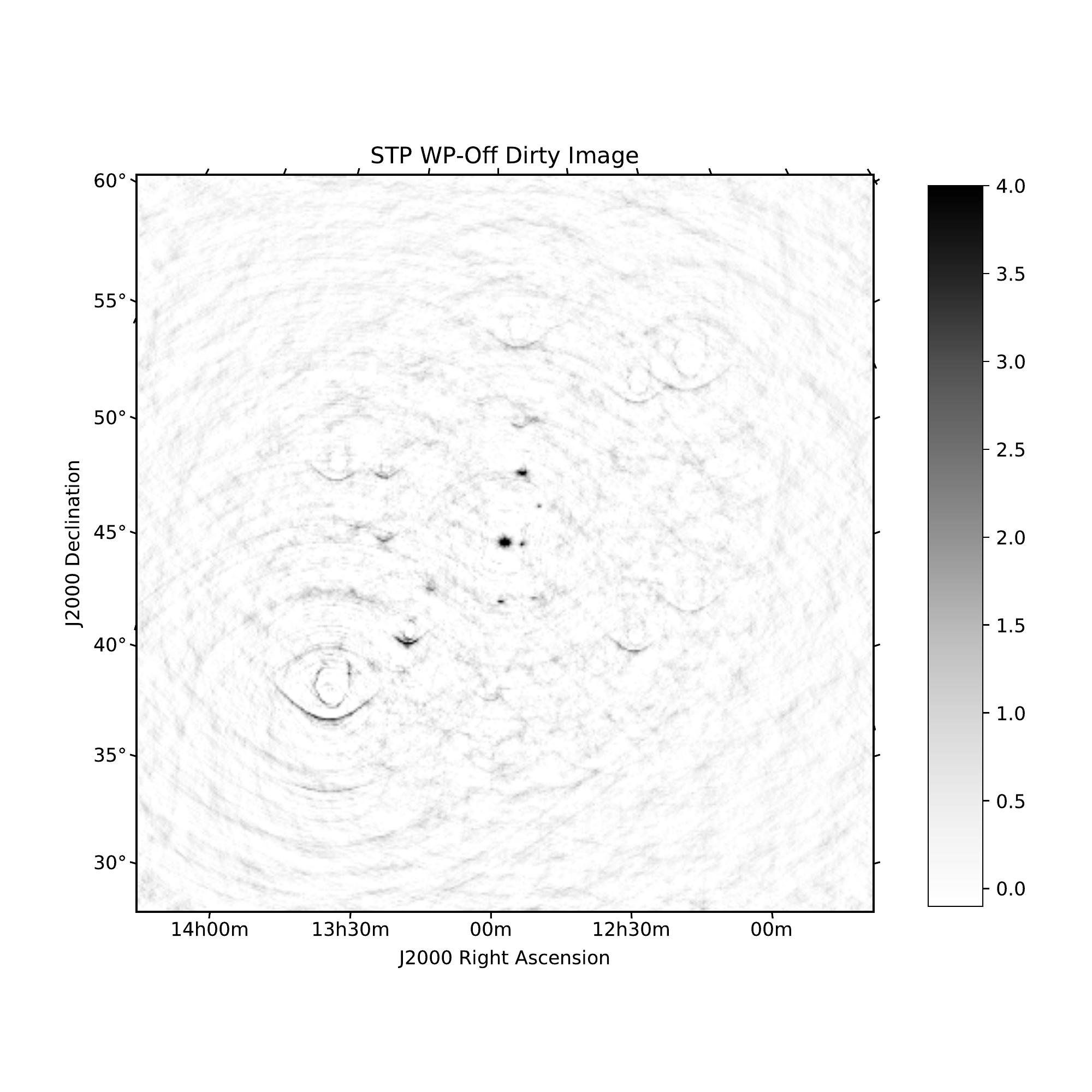}
    \includegraphics[width=0.48\textwidth, clip, trim=0cm 2cm 0cm 1cm]{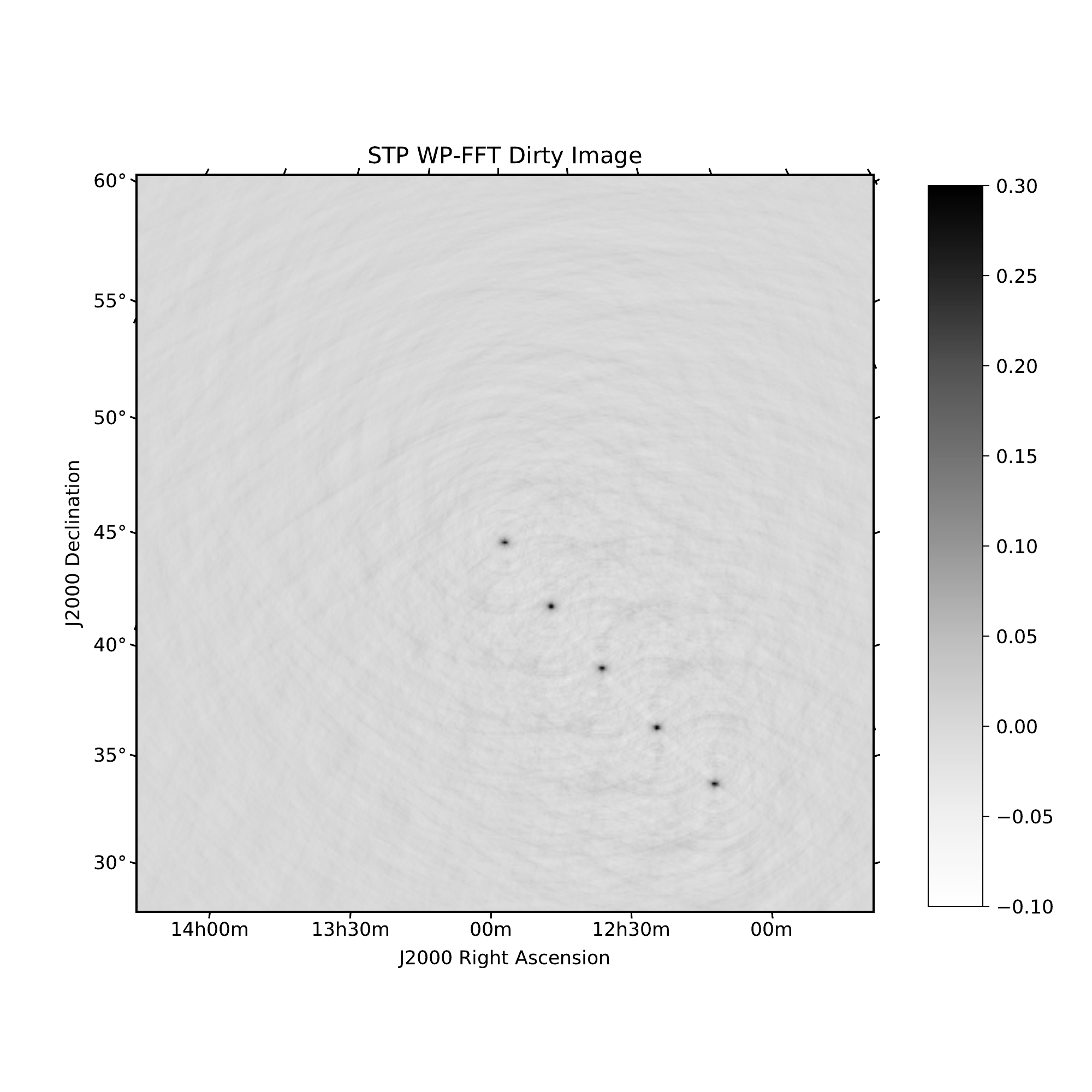}\quad\quad
    \includegraphics[width=0.48\textwidth, clip, trim=0cm 2cm 0cm 1cm]{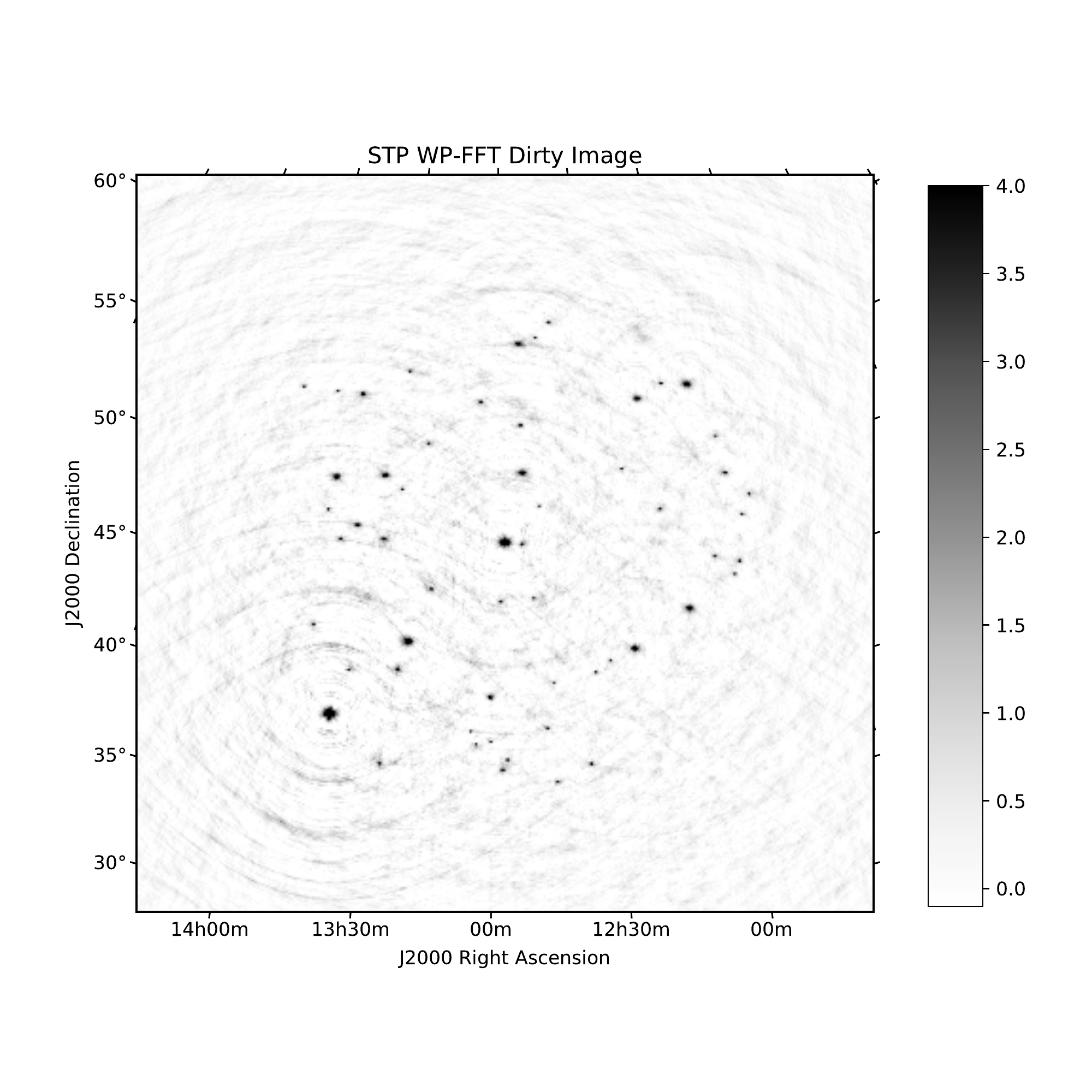}
    \includegraphics[width=0.48\textwidth, clip, trim=0cm 2cm 0cm 1cm]{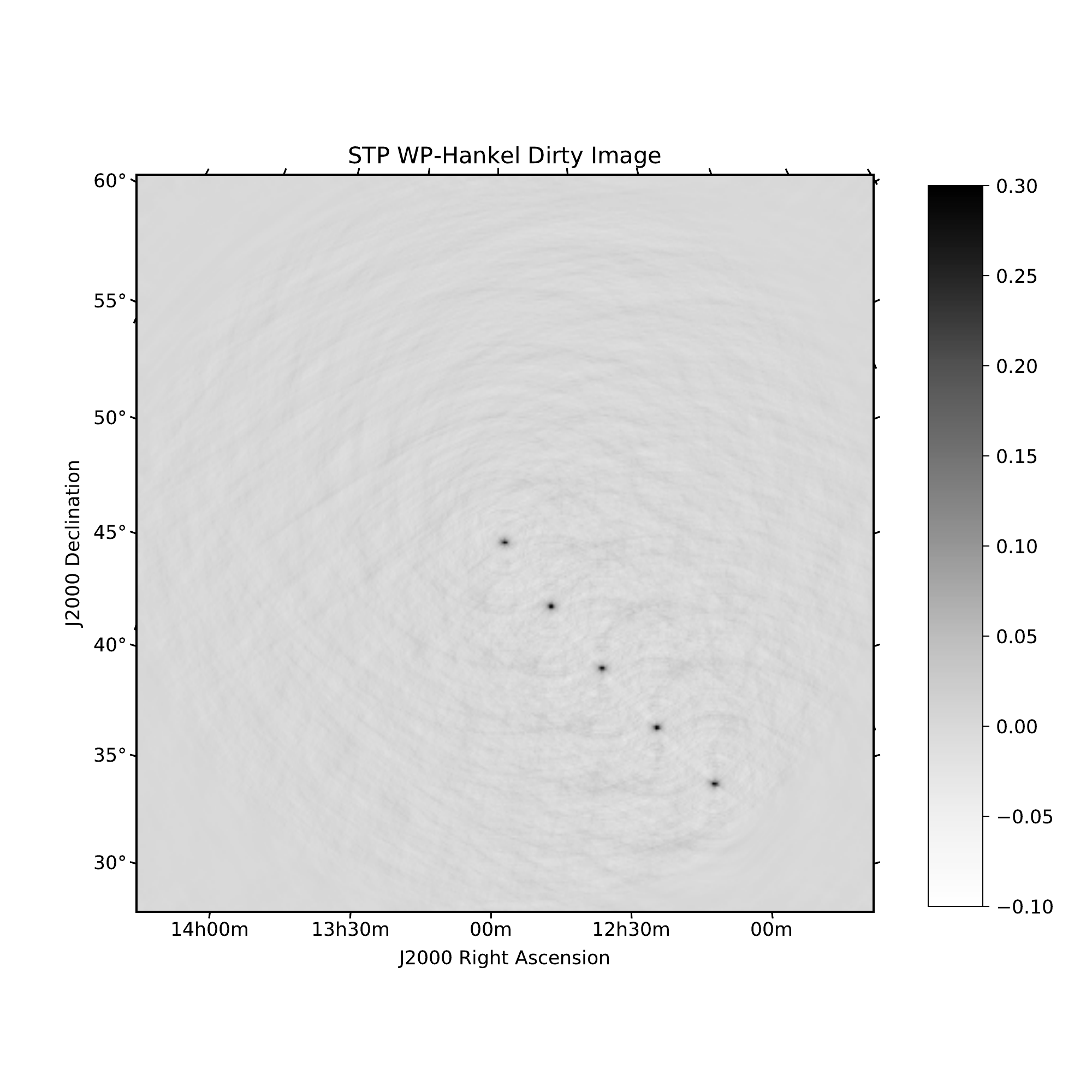}\quad\quad
    \includegraphics[width=0.48\textwidth, clip, trim=0cm 2cm 0cm 1cm]{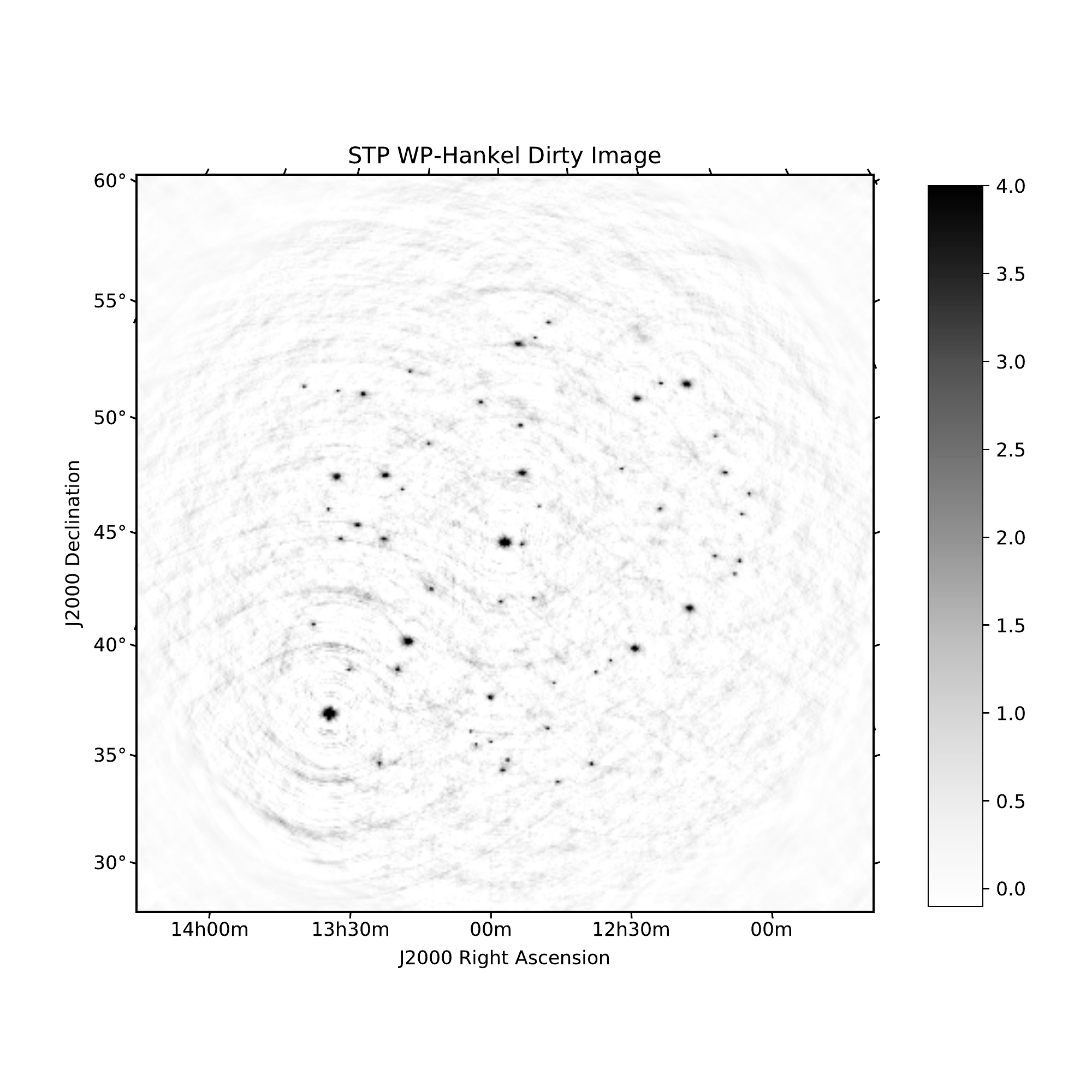}
\caption{Output dirty images produced by STP imager using the 5DIAG (left column) and WENSS (right column) data for three different $w$-projection configurations: \emph{WP-Off} (top row), \emph{WP-FFT} (middle row) and \emph{WP-Hankel} (bottom row).}
\label{fig:output_dirty_images}
\end{figure*}

The output dirty images generated by the three evaluated $w$-projection configurations are shown in Figure \ref{fig:output_dirty_images} for both 5DIAG (left column) and WENSS (right column) measurement sets. From these pictures, the necessity of $w$-projection for accurate generation of wide field images is evident. In the case of the optimised $w$-projection, one may notice (mainly in the WENSS result) that the image corners are attenuated. This is a consequence of the radial convolution kernel generated by interpolation, meaning that the optimised $w$-projection only works within a circular region of the output dirty image with a diameter of image size.
In the case of the simulated test data, the image size and cell size parameters where chosen so that the obtained field-of-view captures all the expected sources with a comfortable margin to the border (padded region).

\subsection{Influence of the number of W-planes}
\label{ssec:wplanes}

\begin{figure*}
\centering
\includegraphics[width=0.48\textwidth]{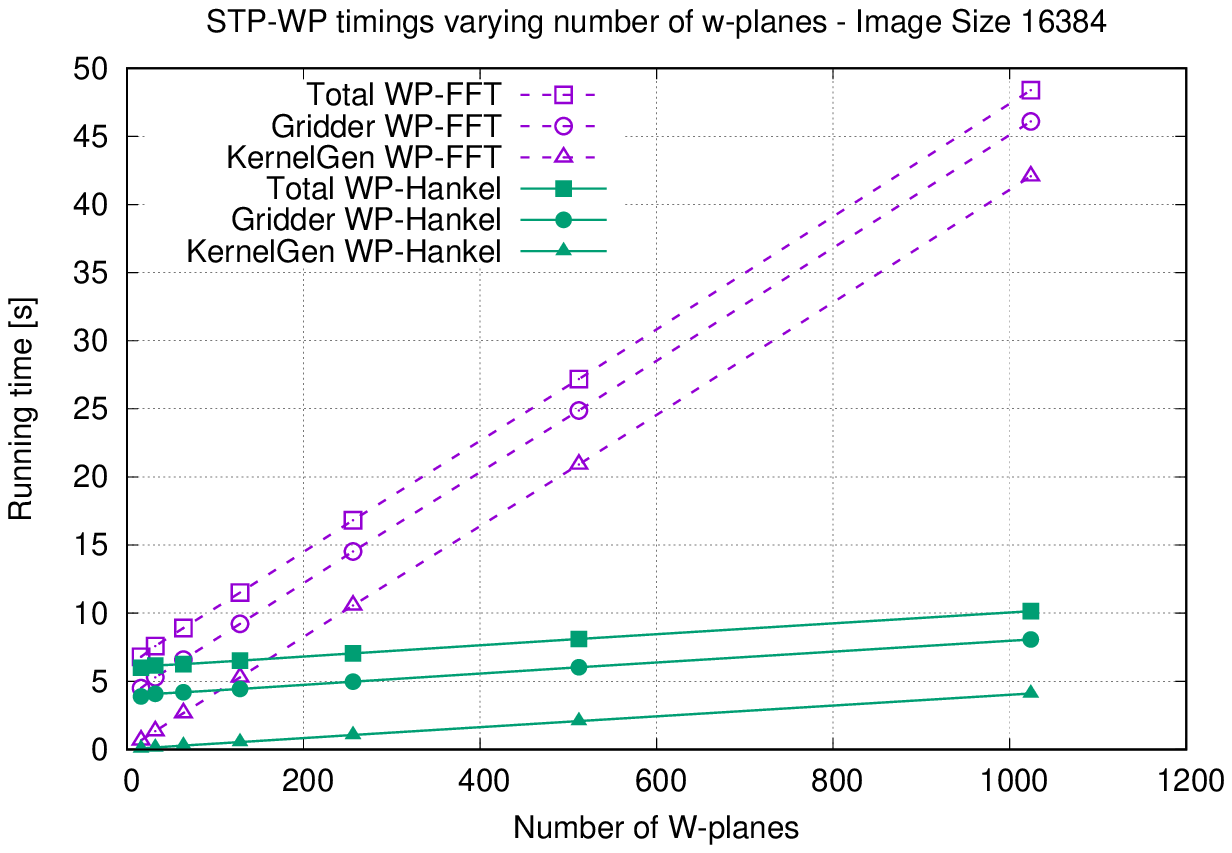}\qquad\includegraphics[width=0.48\textwidth]{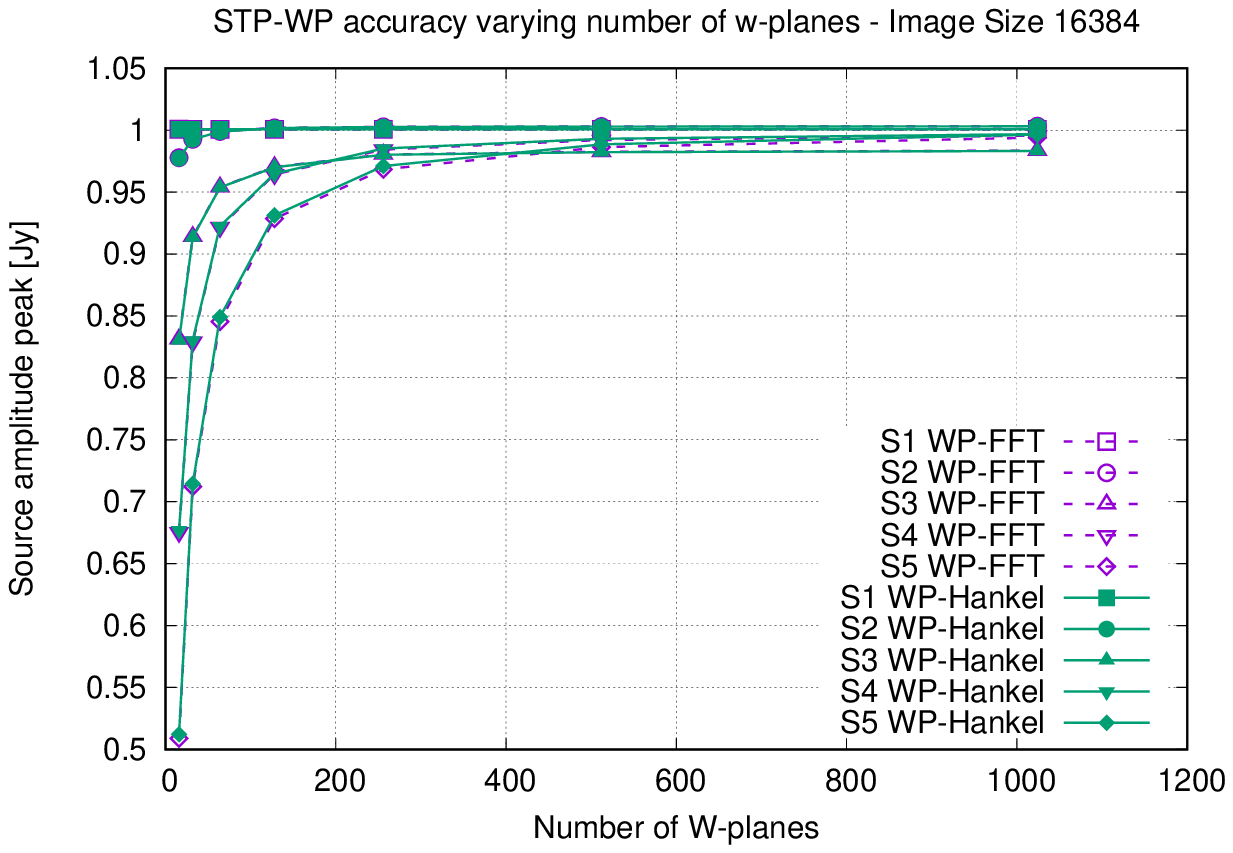}
\caption{Run time results [seconds] (left) and source peak amplitude results [Jansky] (right) of the \emph{WP-FFT} and \emph{WP-Hankel} $w$-projection configurations, using 5DIAG measurement set and IM-L imager configuration (image size $16384 \times 16384$), varying the number of $w$-planes.}
\label{fig:vary_numwplanes_iml}
\end{figure*}

Here we discuss the influence of the number of $w$-planes on the computational performance of the imager.
Figure~\ref{fig:vary_numwplanes_iml} shows the run time results (left image) and the source peak amplitude results (right image) for the \textit{WP-FFT} and \textit{WP-Hankel} configurations as a function of the number of $w$-planes (varying from 16 up to 1024), using the IM-L imager configuration (image size $16384 \times 16384$) and the 5DIAG data. 

In $w$-projection, a distinct convolution kernel needs to be generated to grid the visibility records associated with each $w$-plane. 
As the number of $w$-planes increases, the run time of the convolution kernel generation step will increase in the same proportion. This observation is demonstrated by the results of Figure~\ref{fig:vary_numwplanes_iml}\,(left), where the run time curves increase linearly with the number of $w$-planes. However, the increase rate is significantly different for \emph{WP-FFT} and \emph{WP-Hankel} configurations.
By comparing the \emph{KernelGen} curve of \emph{WP-FFT} and \emph{WP-Hankel}, we can estimate that \emph{WP-Hankel} kernel generation is about 10 times faster than \emph{WP-FFT} in this experiment. Note that this result may depend on the test data, since the generated kernel sizes depend on the actual $w$-plane values, a consequence of the $1\%$ kernel truncation feature.

Regarding the source peak amplitude results in the right image of Figure~\ref{fig:vary_numwplanes_iml}, we observe that the number of $w$-planes is important for the recovered source accuracy. As the number of $w$-planes increases, the source peak amplitude gets closer to the expected strength of 1.0\,Jy. This is most pronounced for the furthest sources, such as S5 (at 12 degrees from the field centre), which reduces the amplitude error from 0.5\,Jy to 0.02\,Jy when the number of $w$-planes changes from 16 to 1024.
For a limited amplitude error, less than 5\%, at least 200 $w$-planes should be used. The benefit of increasing the number of $w$-planes above 512 is quite limited, as the accuracy improvements are minimal and the computational complexity increase is high.

Comparing the source peak amplitude results between \textit{WP-FFT} and \textit{WP-Hankel} configurations, we observe almost no relevant difference, showing that \textit{WP-Hankel} configuration can obtain the same source amplitude accuracy results by using much less computational effort.
This is one of the main advantages of the proposed solution, which scales slowly with the number of $w$-planes, enabling thus more accurate results for the same computational complexity, by using a larger number of $w$-planes.

Another interesting observation in the run time results of Figure~\ref{fig:vary_numwplanes_iml} is the fact that most of the time of the gridding step in \emph{WP-FFT} configuration is used for the convolution kernel generation. The remaining time in the gridding step (given by the difference between curves \emph{Gridder} and \emph{KernelGen}) is mainly used for the gridder convolution operation. This time is similar for both \emph{WP-FFT} and \emph{WP-Hankel}, essentially because the same number of visibility records is used in both experiments. Note that the convolution operation time will vary only with the number of visibility records and this  can be observed in all the experiments presented in this paper. The gain provided by the optimised $w$-projection is visible only in the kernel generation step.

From Figure~\ref{fig:vary_numwplanes_iml} we can also observe that most of the imager time (\emph{Total} curve) is used by the gridding step. The observed difference between the \emph{Total} and \emph{Gridder} curves is the time used by the Fourier inversion of the gridded data (\emph{i.e. FFT}) that generates the dirty image. The complexity of this FFT depends on the image size used. For large image sizes its complexity can be larger than the gridder complexity. In the following, we analyse the impact of the image size on $w$-projection.

\subsection{Influence of the image size}
\label{ssec:image_size}

\begin{figure*}
\centering
\includegraphics[width=0.48\textwidth]{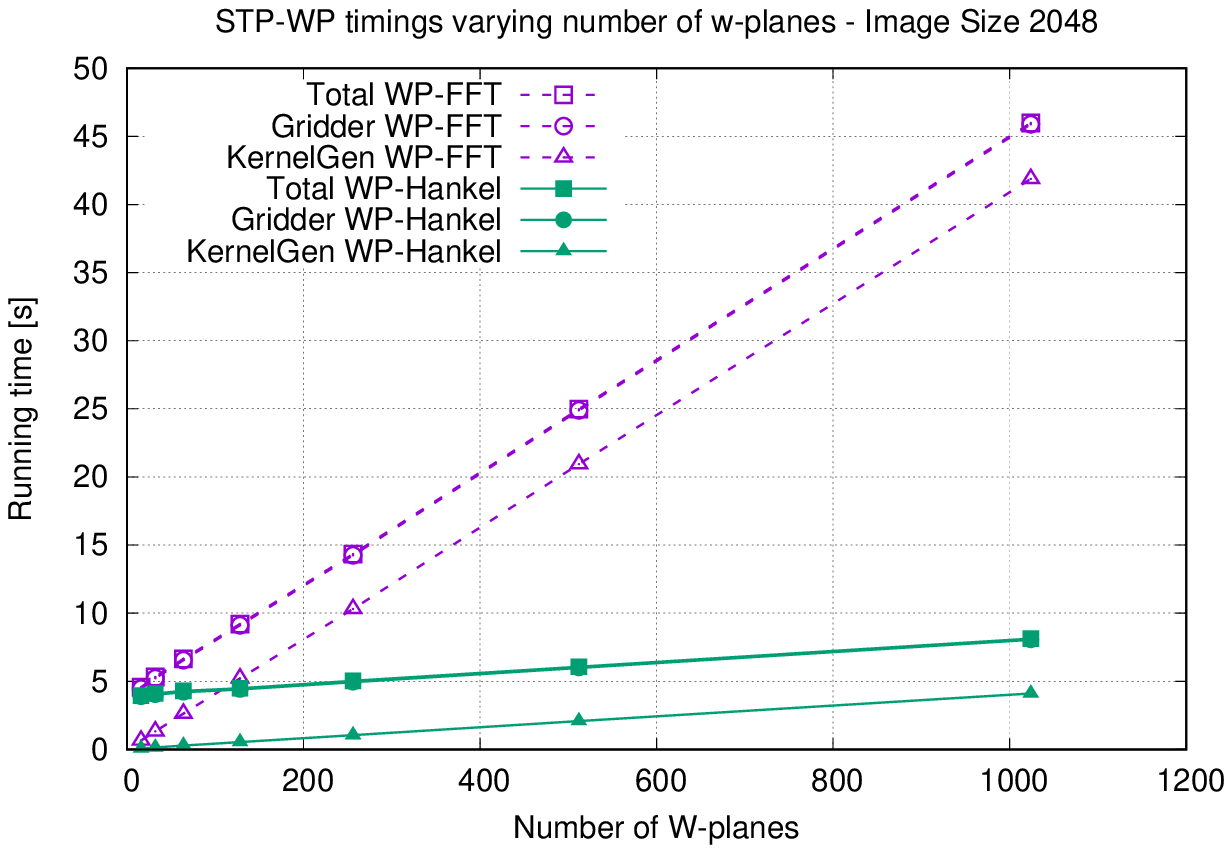}\qquad\includegraphics[width=0.48\textwidth]{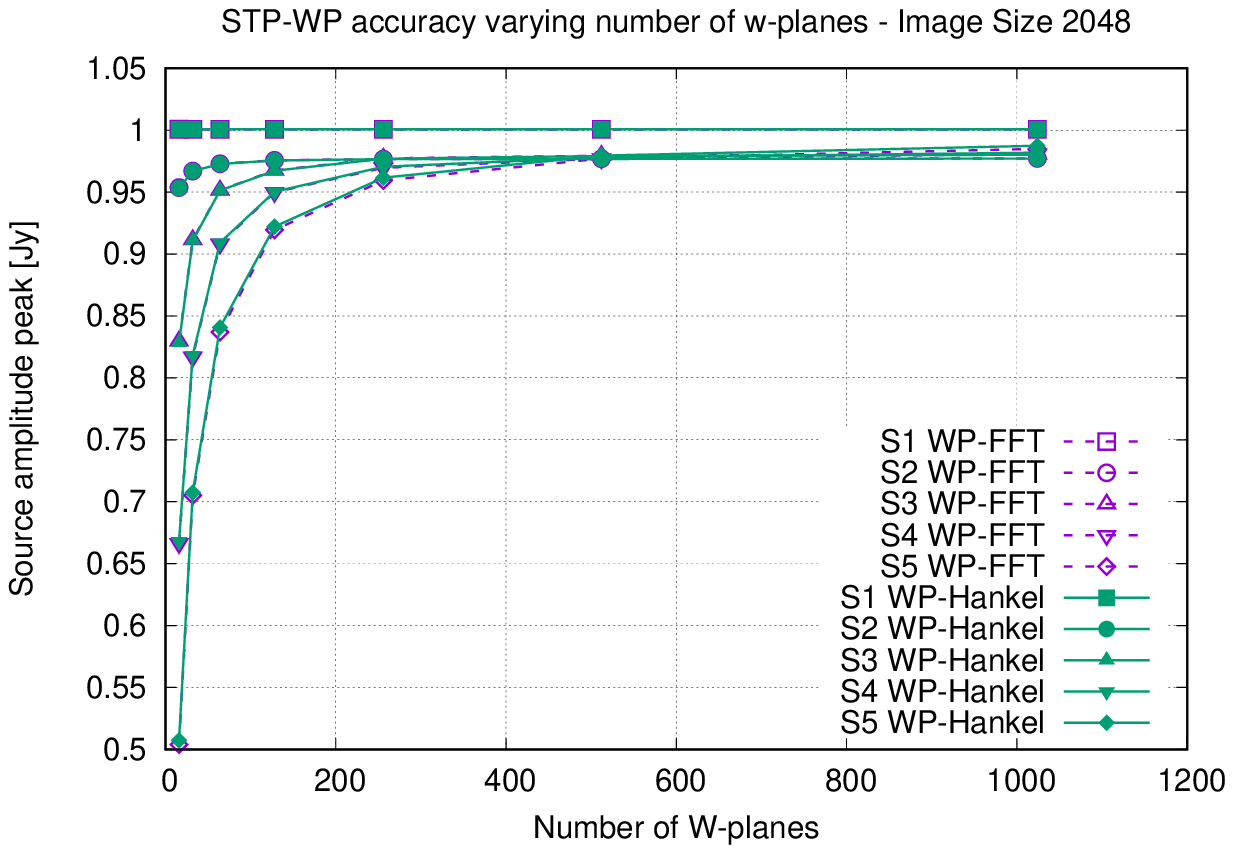}
\caption{Run time results [seconds] (left) and source peak amplitude results [Jansky] (right) of the \emph{WP-FFT} and \emph{WP-Hankel} $w$-projection configurations, using 5DIAG measurement set and IM-S imager configuration (image size $2048 \times 2048$), varying the number of $w$-planes.}
\label{fig:vary_numwplanes_ims}
\end{figure*}

To show the influence of the image size in $w$-projection performance, we repeated the previous experiments for 5DIAG data using a smaller image size, as specified by IM-S imager configuration (image size $2048 \times 2048$). Note that the IM-S configuration also changes the cell size, so that the field-of-view is kept constant.

The results obtained are presented in Figure~\ref{fig:vary_numwplanes_ims}. As can be observed, the run times of the gridding and kernel generation steps are quite similar to the ones of Figure~\ref{fig:vary_numwplanes_iml} presented in previous subsection for image size $16384\times 16384$. The main noticeable difference is the imager \emph{Total} run time curve, which in Figure~\ref{fig:vary_numwplanes_ims} is almost overlapping the \emph{Gridder} curve. 
This result can be explained by the fact that the FFT of the gridded data with size $2048 \times 2048$ presents a quite small computational complexity when compared to the gridding step. Thus, the imager run time is mostly the gridding step run time. Regarding the source peak amplitude results, the same conclusions as the ones presented in the previous subsection for the $16384\times 16384$ image size are drawn.
Given the unnoticeable influence of the image size in the gridder run time, the remaining experiments of the paper use only the IM-L imager configuration.

\subsection{Influence of convolution kernel size}

\begin{figure*}
\centering
\includegraphics[width=0.48\textwidth]{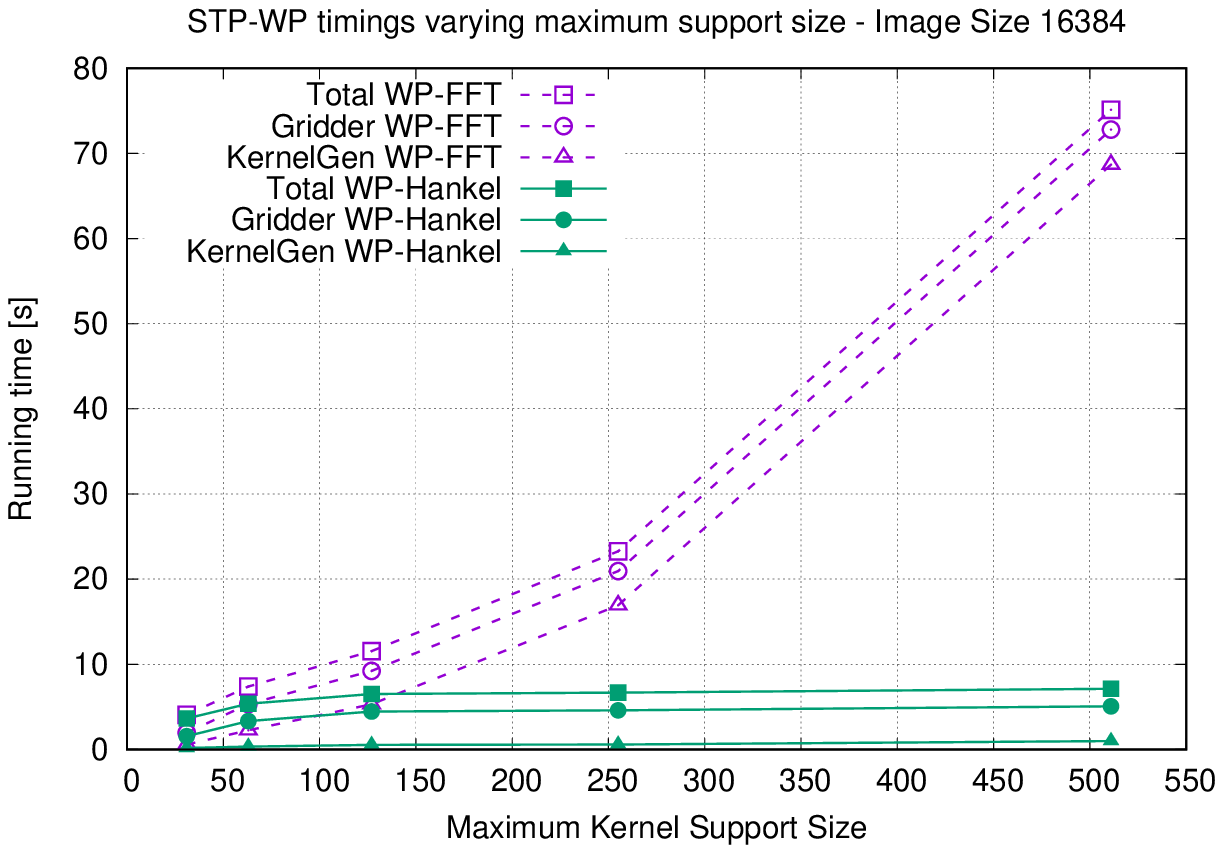}\qquad\includegraphics[width=0.48\textwidth]{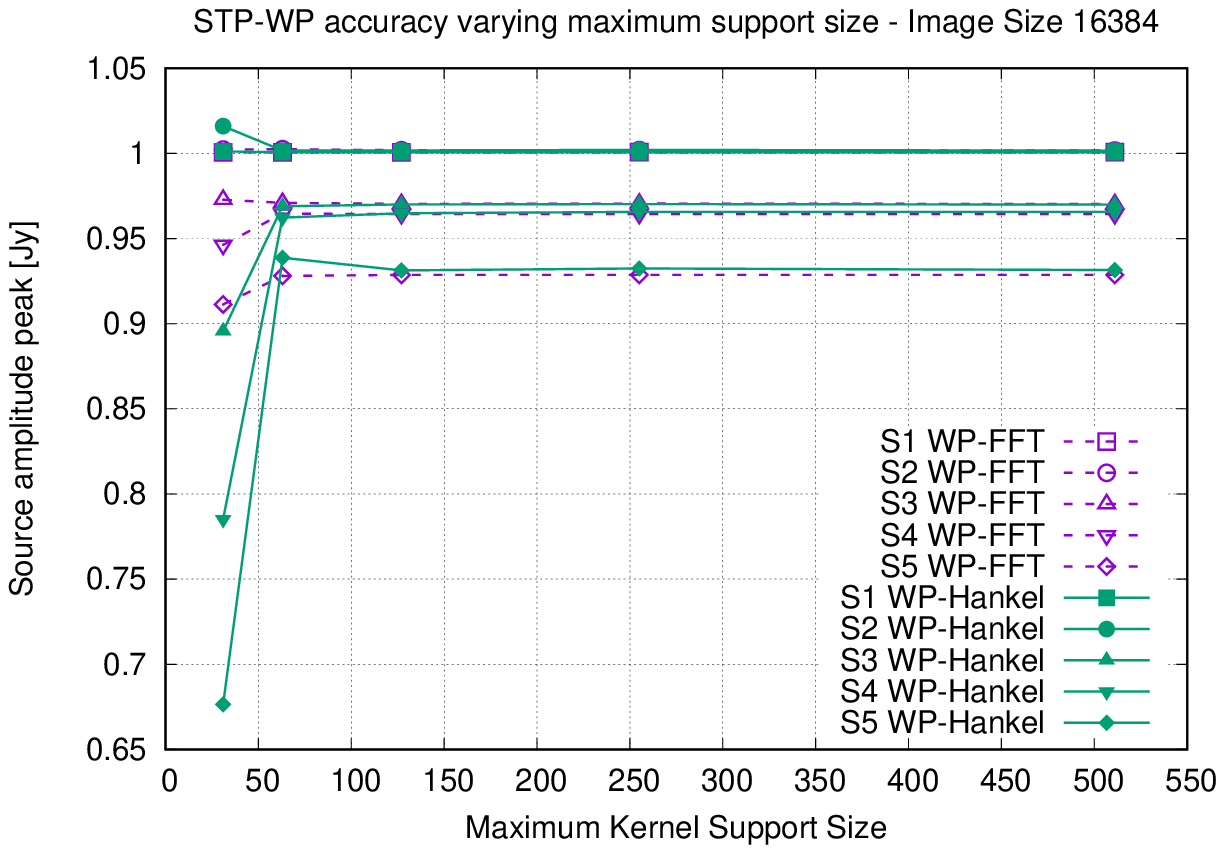}
\caption{Run time results [seconds] (left) and source peak amplitude results [Jansky] (right) of the \emph{WP-FFT} and \emph{WP-Hankel} $w$-projection configurations, using 5DIAG measurement set and IM-L imager configuration (image size $16384 \times 16384$), varying the maximum support size of the convolution kernel.}
\label{fig:vary_maxconvsize_iml}
\end{figure*}

The maximum allowed support size for the convolution kernel is an input parameter of the imager algorithm. It is used to determine the workspace size when the undersampling optimisation is enabled, instead of using the image size. The workspace size shall be a power of two value immediately above the maximum kernel size given by $2\times \text{support} + 1$. Note that we use a power of two workspace size to enable the use of pre-optimised FFTW plans, which are more efficient than runtime plan optimisation. When oversampling is used, the workspace is further increased by using zero padding by an amount equal to the oversampling ratio (typically 8 times).

Figure~\ref{fig:vary_maxconvsize_iml} presents the run time results (left image) and the source peak amplitude results (right image) for the \textit{WP-FFT} and \textit{WP-Hankel} configurations as a function of the maximum kernel support, using the IM-L imager configuration (image size $16384 \times 16384$) and the 5DIAG data. To make the maximum kernel size as close as possible from a power of two value (and thus the workspace size), we used the following support values: 31, 63, 127, 255 and 511, which corresponds to the workspace sizes: 64, 128, 256, 512 and 1024, respectively. 

The experimental results show that the kernel generation complexity in \textit{WP-Hankel} configuration increases slightly with the support size, while it presents an exponential grow in \textit{WP-FFT} configuration. This difference can be explained by the fact that most computational complexity of the kernel generation step in \textit{WP-FFT} configuration is associated to the two-dimensional FFT with the size of the workspace zero padded by the oversampling factor of 8. Thus, due to the oversampling feature, the sizes of the 2D-FFT performed for the curve points shown in Figure~\ref{fig:vary_maxconvsize_iml} are: $512\times 512$, $1024 \times 1024$, $2048 \times 2048$, $4096 \times 4096$ and $8192\times 8192$. The equivalent one-dimensional sizes are used for the Hankel transform in \textit{WP-Hankel} configuration.

The kernel generation results of \textit{WP-FFT} and \textit{WP-Hankel} configurations can thus be explained mainly by the asymptotic complexities of the 2D-FFT and 1D-FFT (used for Hankel transform), which are $\mathcal{O}(N^2\log{N^2})$ and $\mathcal{O}(N\log{N})$, respectively. To be precise, the complexity of Hankel transform approach must account also for the kernel projection and kernel interpolation steps. Furthermore, both \textit{WP-FFT} and \textit{WP-Hankel} need to build the oversampled kernel cache as part of the kernel generation step. Regarding the complexity of the gridder convolution operation, no difference is noticeable between both configurations (see the difference between \emph{Gridder} and \emph{KernelGen} curves), as expected.

Although the final convolution kernel size is determined by the kernel truncation feature, the use of a large enough workspace is required to enable wide convolution kernels which tend to occur when $w$-term is large. 
For instance, in these experiments it was observed that the largest convolution kernel size generated with 1\% kernel truncation was 255. Thus, in order to generate kernels up to the 255 size, the support size can not be less than 127.
The use of optimised $w$-projection is thus advantageous as it can handle larger workspaces without a noticeable impact on the computational complexity of the algorithm. 

Regarding the source peak amplitude results of Figure~\ref{fig:vary_maxconvsize_iml}, we observe that both \textit{WP-FFT} and \textit{WP-Hankel} configurations present similar and well-behaved results for support sizes above 127. However, for small support values, a significant amplitude error is created because the workspace size is not enough to represent the full convolution kernel with 1\% truncation. Such situations should not occur and they can be prevented by setting a large enough value for the maximum kernel support.

\subsection{Influence of the kernel truncation percentage}

\begin{figure*}
\centering
\includegraphics[width=0.48\textwidth]{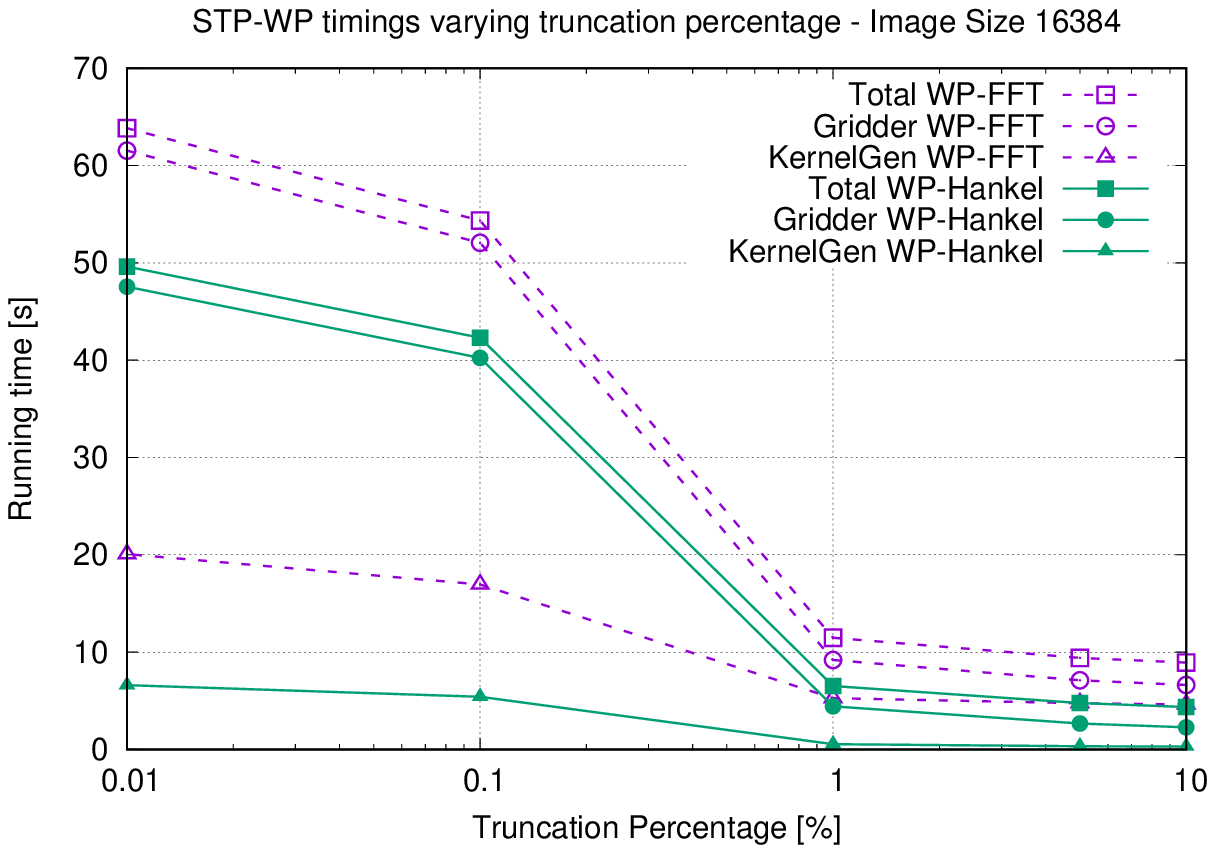}\qquad\includegraphics[width=0.48\textwidth]{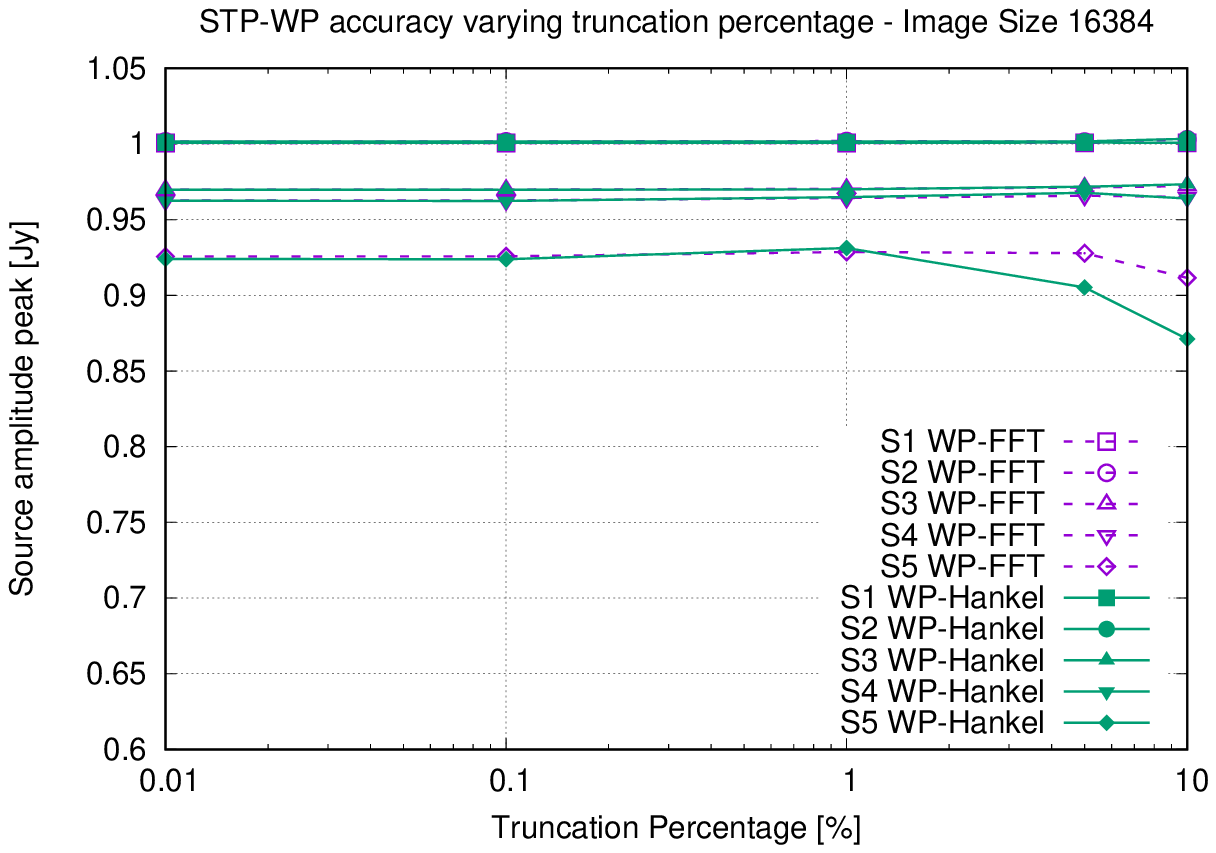}
\caption{Run time results [seconds] (left) and source peak amplitude results [Jansky] (right) of the \emph{WP-FFT} and \emph{WP-Hankel} $w$-projection configurations, using 5DIAG measurement set and IM-L imager configuration (image size $16384 \times 16384$), varying the kernel truncation percentage.}
\label{fig:vary_truncation_iml}
\end{figure*}

Here, we analyse the influence of the convolution kernel truncation percentage on the imager computational performance and output source amplitude. The kernel truncation is applied to reduce the gridder computational effort at the cost of some accuracy loss in the output source amplitude. In these tests, we vary kernel truncation percentage between 0.01\% and 10\%, but typically 1\% is used.
The results using \textit{WP-FFT} and \textit{WP-Hankel} configurations for the 5DIAG data and IM-L imager configuration (image size $16384 \times 16384$) are presented in Figure~\ref{fig:vary_truncation_iml}.

From the presented results, we can observe that decreasing the kernel truncation percentage tends to increase the gridding step and consequently the imager run time. A smaller truncation percentage originates larger convolution kernels which significantly increase the gridder convolution time for both \textit{WP-FFT} and \textit{WP-Hankel} configurations (see the difference between \emph{Gridder} and \emph{KernelGen} curves). It also increases the kernel generation time, mostly due to the kernel cache building step, which needs to store larger kernel arrays.
For the smaller kernel truncation value ($0.01\%$), the results of Figure~\ref{fig:vary_truncation_iml} seem to saturate, because the most kernels are not being truncated, i.e. their size is limited by the maximum support of 127 pixels (with the maximum size of a convolution kernel being 255).

In the case of the \textit{WP-Hankel}, the reduced kernel truncation percentage also affects the interpolation step negatively, as it depends on the truncated kernel size. We can observe in Figure~\ref{fig:vary_truncation_iml}, that the kernel generation speedup of \textit{WP-Hankel} for the kernel truncation of $0.01\%$ is about 3 times, rather than 10 times, as discussed in \S~\ref{ssec:wplanes} for 1\% kernel truncation. However, such small kernel truncation values shall not be used, as they do not provide noticeable improvement in the source amplitude results. When increasing the truncation percentage above 1\%, large amplitude errors can arise, as observed in Figure~\ref{fig:vary_truncation_iml}.

\subsection{Influence of the oversampling ratio}

\begin{figure*}
\centering
\includegraphics[width=0.48\textwidth]{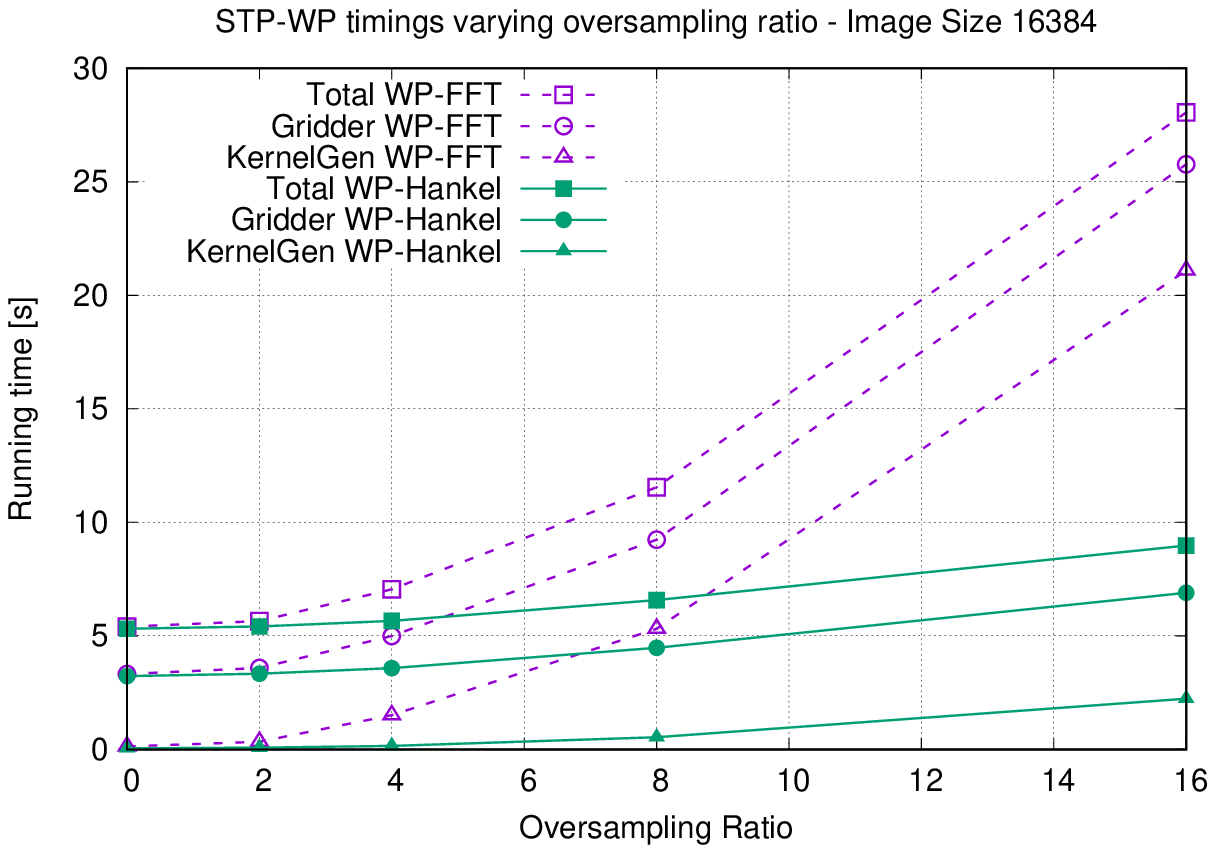}\qquad\includegraphics[width=0.48\textwidth]{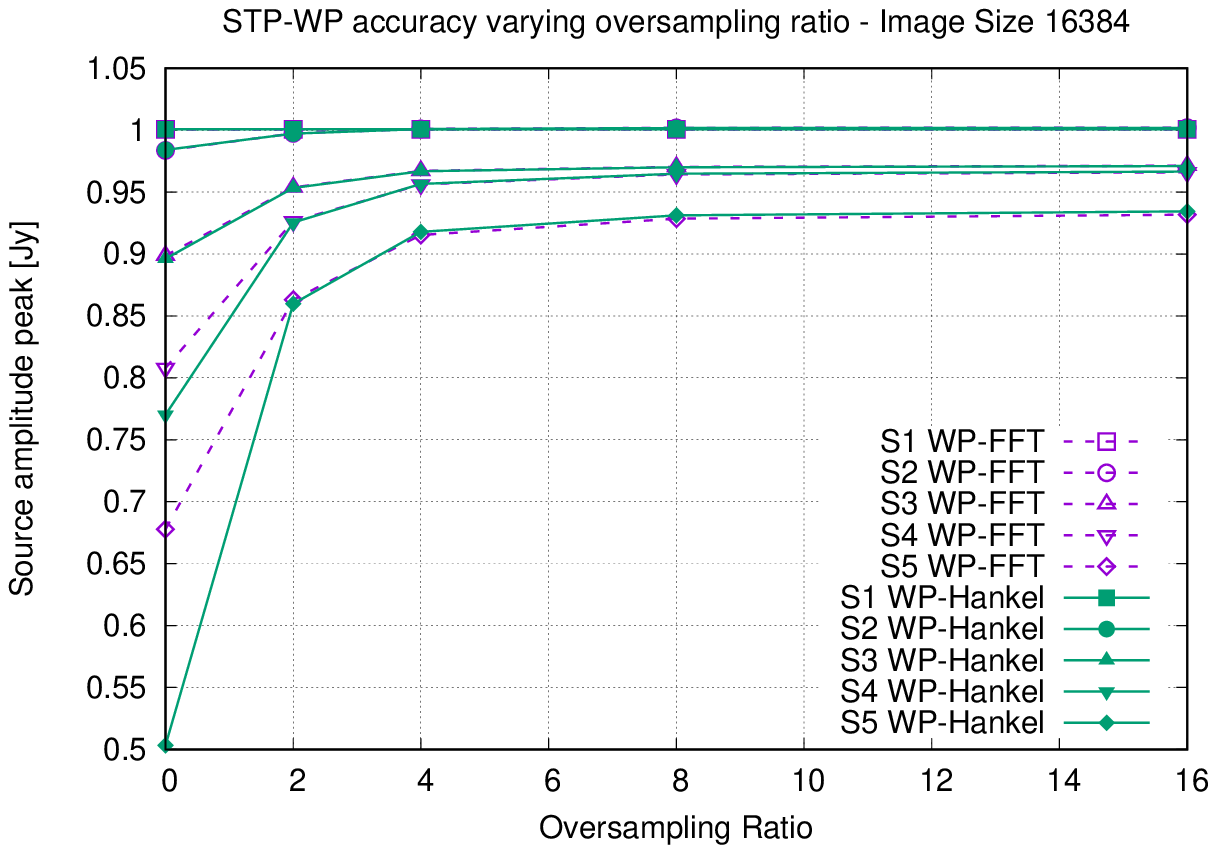}
\caption{Run time results [seconds] (left) and source peak amplitude results [Jansky] (right) of the \emph{WP-FFT} and \emph{WP-Hankel} $w$-projection configurations, using 5DIAG measurement set and IM-L imager configuration (image size $16384 \times 16384$), varying the oversampling ratio.}
\label{fig:vary_oversampling_iml}
\end{figure*}

The influence of the oversampling ratio in the imager performance is shown in Figure~\ref{fig:vary_oversampling_iml} for the 5DIAG data and IM-L imager configuration (image size $16384 \times 16384$). It can be observed that the oversampling ratio has a high impact on the run time of the kernel generation step. This result is explained by the fact that oversampling is achieved by zero padding the workspace area resulting in much larger FFT sizes and consequently slower executions. The impact on the computational complexity is significantly larger on the \textit{WP-FFT} because the FFT for convolution kernel generation is two dimensional. 

The peak amplitude results of the detected sources prove that the use of oversampling is of utmost importance. Significant amplitude errors are obtained when oversampling is disabled, \emph{i.e.} when it is zero. Also, using oversampling values above 8 seems not to be worth as there are no visible improvements in the amplitude results. The use of oversampling is more important for the \textit{WP-Hankel}, because it helps in reducing the errors in the kernel interpolation step. This can be seen in Figure~\ref{fig:vary_oversampling_iml} by the fact that the lack of oversampling causes larger source amplitude errors on the \textit{WP-Hankel} configuration.
Given that oversampling is required for a reasonable output accuracy performance, this should not be an issue of the \textit{WP-Hankel} approach.

\subsection{Influence of the interpolation type}

\begin{figure}
\centering
\includegraphics[width=0.48\textwidth]{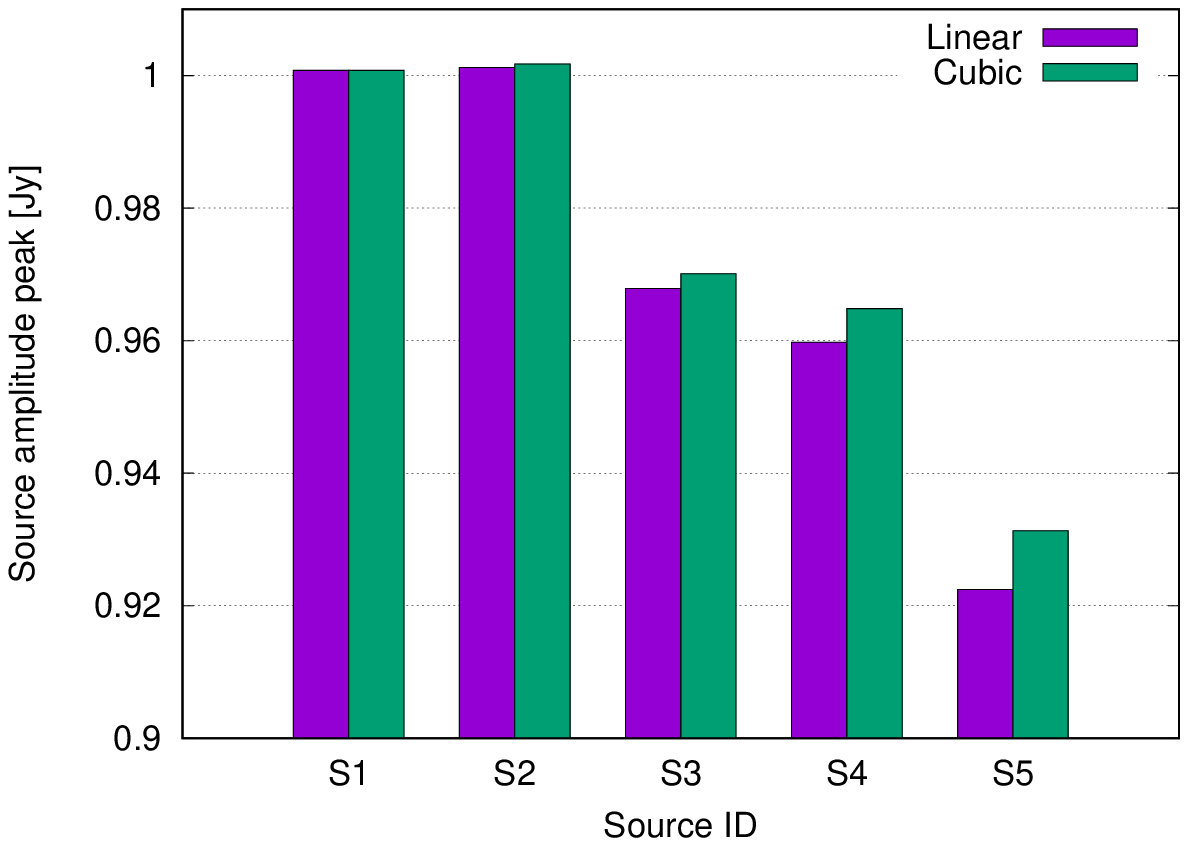}
\caption{Source peak amplitude results [Jansky] of the \emph{WP-Hankel} configuration with linear and cubic interpolation methods, using the 5DIAG measurement set and IM-L imager configuration (image size $16384 \times 16384$).}
\label{fig:vary_interpolation_hankel_accuracy}
\end{figure}

The use of Hankel transform optimisation requires 2D-kernel interpolation from the 1D transformed array. In Figure~\ref{fig:vary_interpolation_hankel_accuracy} we evaluate the imager performance of linear and cubic interpolation techniques when using the IM-L imager configuration (image size $16384 \times 16384$) and \emph{WP-Hankel} configuration with the 5DIAG data.
In terms of the execution performance, it was observed that linear interpolation provides a slight run time reduction of about 0.05 seconds in the total run time of 6.479 seconds. On the other hand, the peak amplitude of each source varies according to Figure~\ref{fig:vary_interpolation_hankel_accuracy} (right image). From these results we may conclude that kernel interpolation has quite small influence in the amplitude error, increasing slightly by a few thousandths of Jansky. However, given the small impact on the computational complexity we opt to use cubic interpolation by default.

\subsection{Multi-threaded performance}

\begin{table}
\centering
\caption{Single- and multi-threaded run times of the imager (in seconds) for the 5DIAG data, using the IM-L imager configuration (image size $16384 \times 16384$) and three distinct $w$-projection configurations: \emph{WP-Off}, \emph{WP-Hankel} and \emph{WP-FFT}.}
\label{tab:single_multi_bench}
\begin{tabular}{|l|c|c|c|}
\hline
\textbf{Config.} & \textbf{Singlethread} & \textbf{Multithread} & \textbf{Parallel} \\
 & \textbf{{[}s{]}} & \textbf{{[}s{]}} & \textbf{Eff.} \\ \hline
WP-Off     & 3.619    &  2.099    & 43\%    \\ \hline
WP-FFT     & 27.688  &  11.514    & 60\%     \\ \hline
WP-Hankel  & 18.179   &  6.479    & 70\%   \\ \hline
\end{tabular}
\end{table}

To evaluate the multi-threading gains of the proposed implementation, we run the imager in both single- and multi- threaded modes and we computed the parallel efficiency using:
\begin{equation}
\eta = \frac{\text{speedup}}{N_{cores}}\times 100 \;\;,
\end{equation}
where \emph{speedup} is the ratio between the single-threaded and multi-threaded run times, and $N_{cores}$ is the number of physical threads of the CPU, which for the case of the machine used in these tests, without using hyper-threading, is given by $N_{cores}=4$.

For single-threaded execution we used the \emph{taskset} tool available in Linux OS to restrict the number of usable CPU cores to one. The results obtained for single- and the multi-threaded executions are shown in Table \ref{tab:single_multi_bench}, for the \textit{WP-FFT}, \textit{WP-Hankel} and \textit{WP-Off} configurations.
We can observe that the tests present a parallel efficiency between 43\% and 70\%. In fact, several technical issues prevent an ideal parallel efficiency, such as thread synchronisation, thread creation overhead, short non-parallelisable tasks, among others.

The use of $w$-projection tends to increase parallel efficiency essentially due to the larger kernel sizes used in the convolutional gridder. These kernels increase the convolution complexity which takes longer in multi-threaded execution improving thus the parallel efficiency. In fact, we observed that the gridding step itself presents a parallel efficiency above 90\% in \emph{WP-Hankel} configuration. The main reason for the lower total parallel efficiency of 70\% is the low multi-threading performance of the FFT calls. This is also the justification for the higher parallel efficiency of \emph{WP-Hankel} when compared with \emph{WP-FFT}, because the latter uses more time performing FFT for the convolution kernel generation.

\section{Comparison to CASApy software}
\label{sec:comparisonCasapy}

\begin{figure}
\centering
\includegraphics[width=0.48\textwidth]{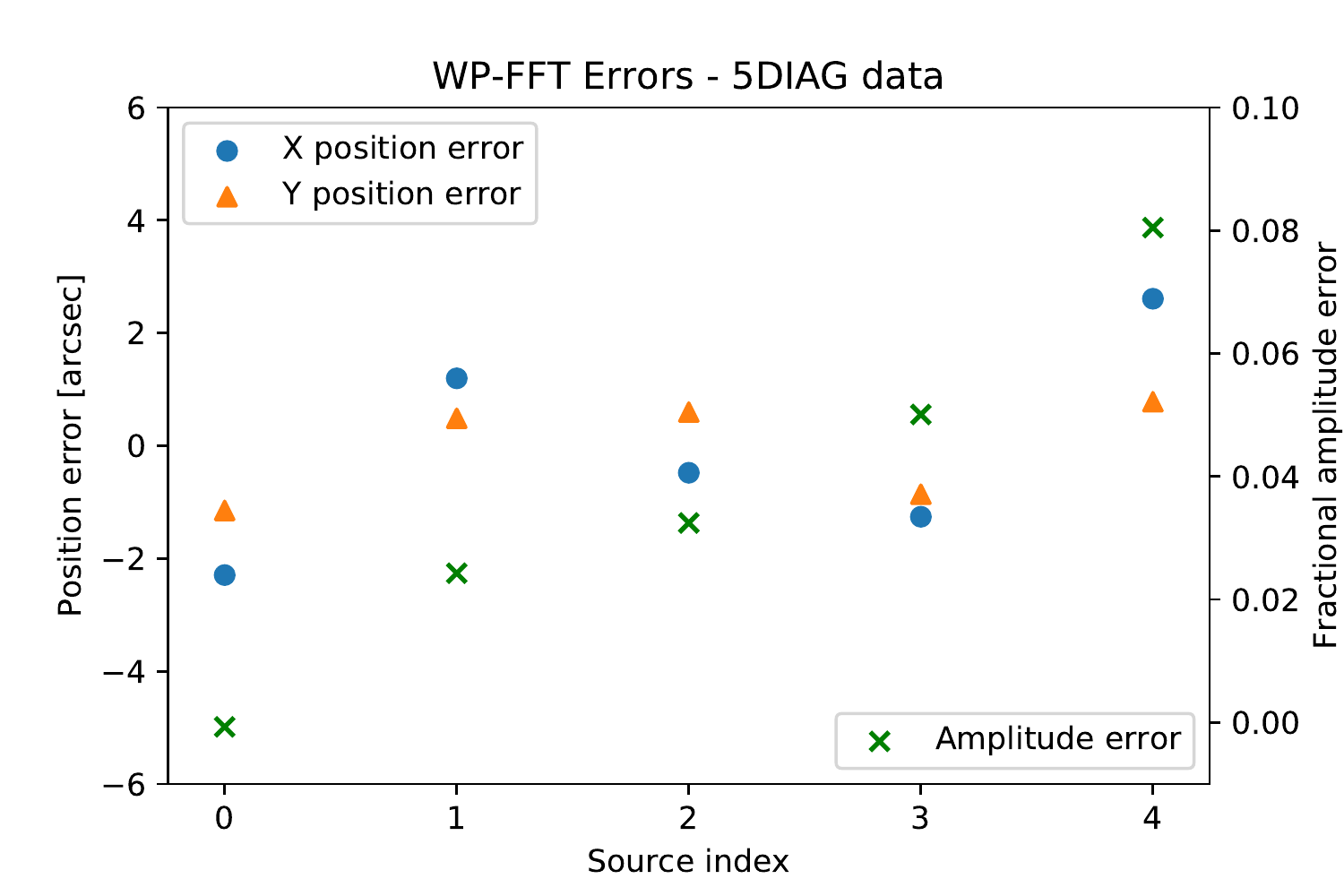}
\includegraphics[width=0.48\textwidth]{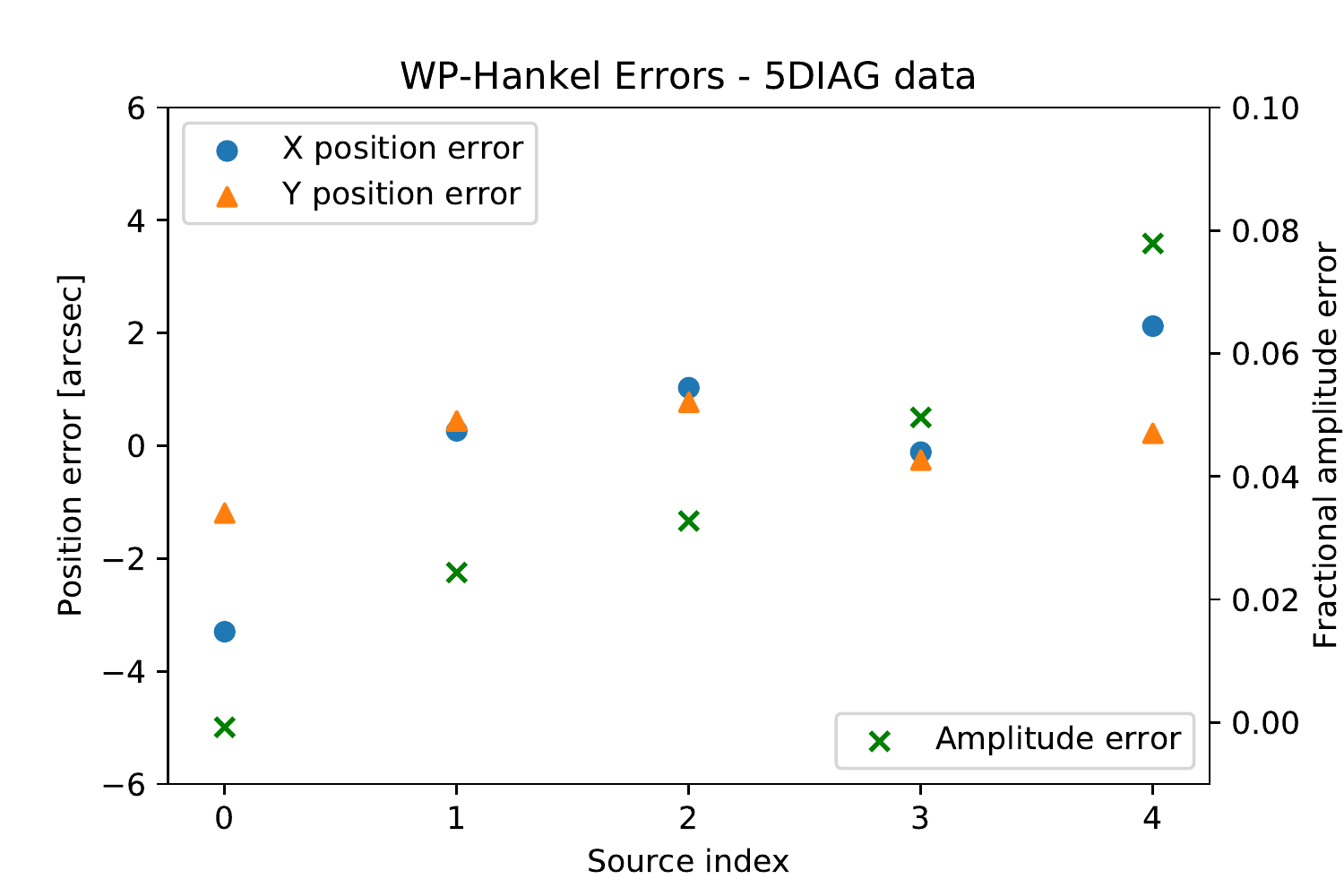}
\includegraphics[width=0.48\textwidth]{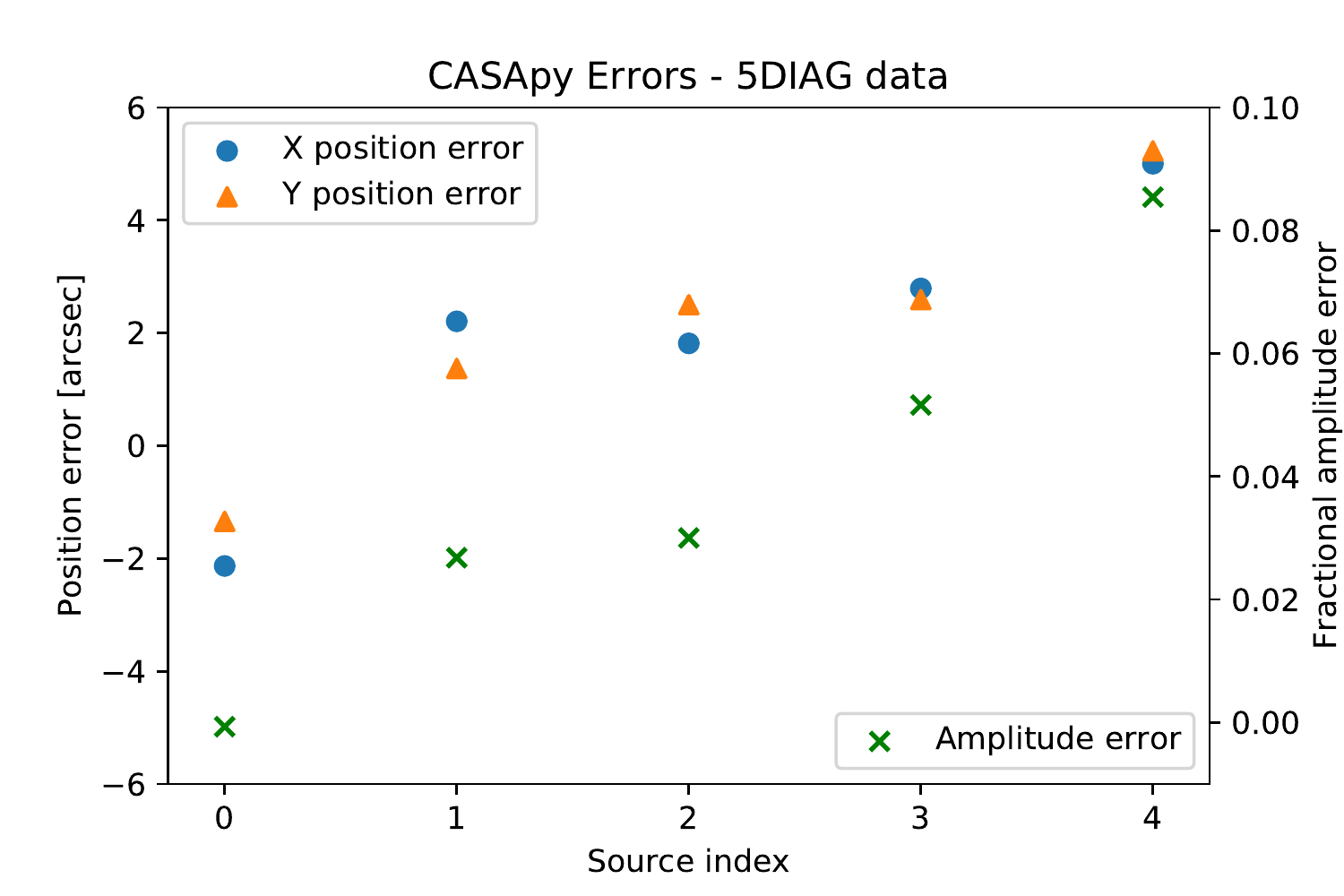}
\caption{Source peak amplitude errors [Jansky] and source position errors [arcsec] using \emph{WP-FFT} (top), \emph{WP-Hankel} (middle) and CASApy (bottom), the 5DIAG measurement set and IM-S imager configuration (image size $2048 \times 2048$).}
\label{fig:wpfft_errors_5diag}
\end{figure}

\begin{figure*}
\centering
\includegraphics[width=\textwidth]{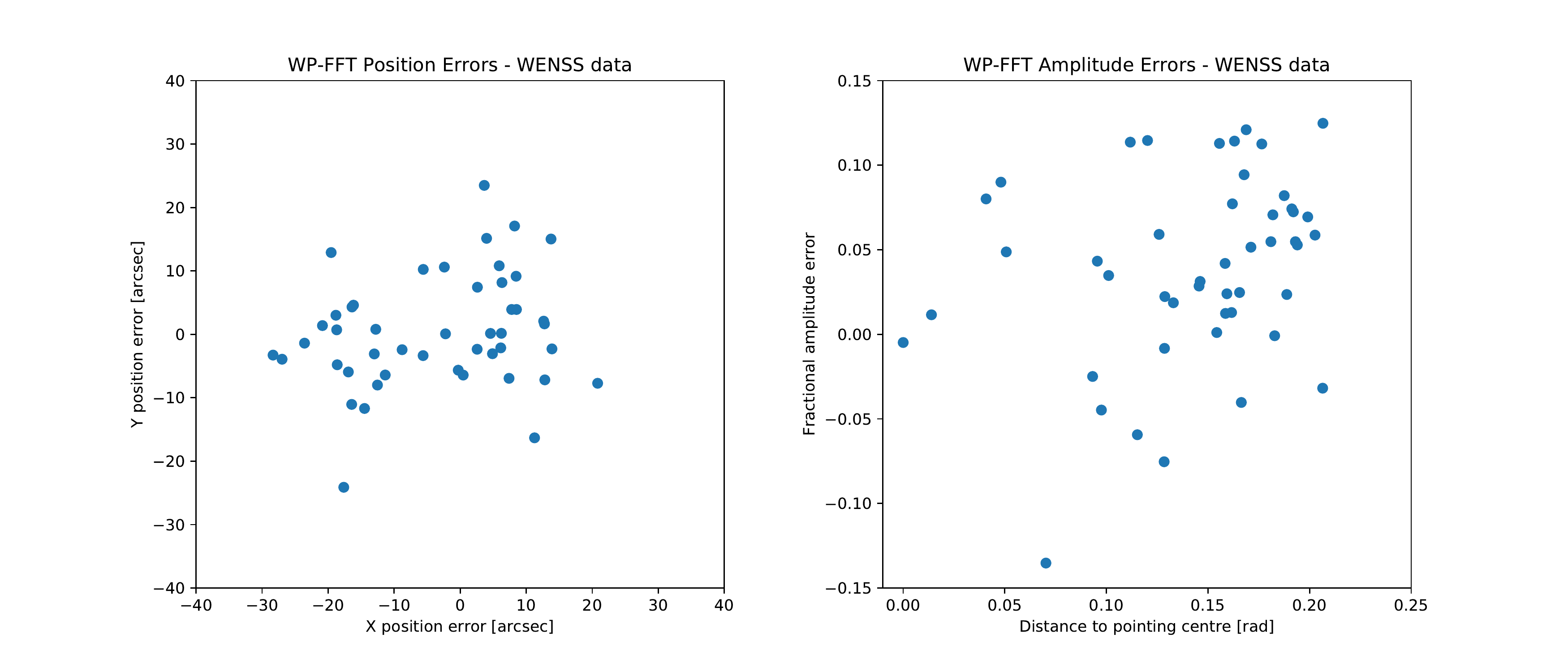}
\caption{Source position errors [arcsec] (left) and source peak amplitude errors [Jansky] (right) using the \emph{WP-FFT} configuration, the 5DIAG measurement set and IM-S imager configuration (image size $2048 \times 2048$).}
\label{fig:wpfft_errors_wenss}
\end{figure*}

\begin{figure*}
\centering
\includegraphics[width=\textwidth]{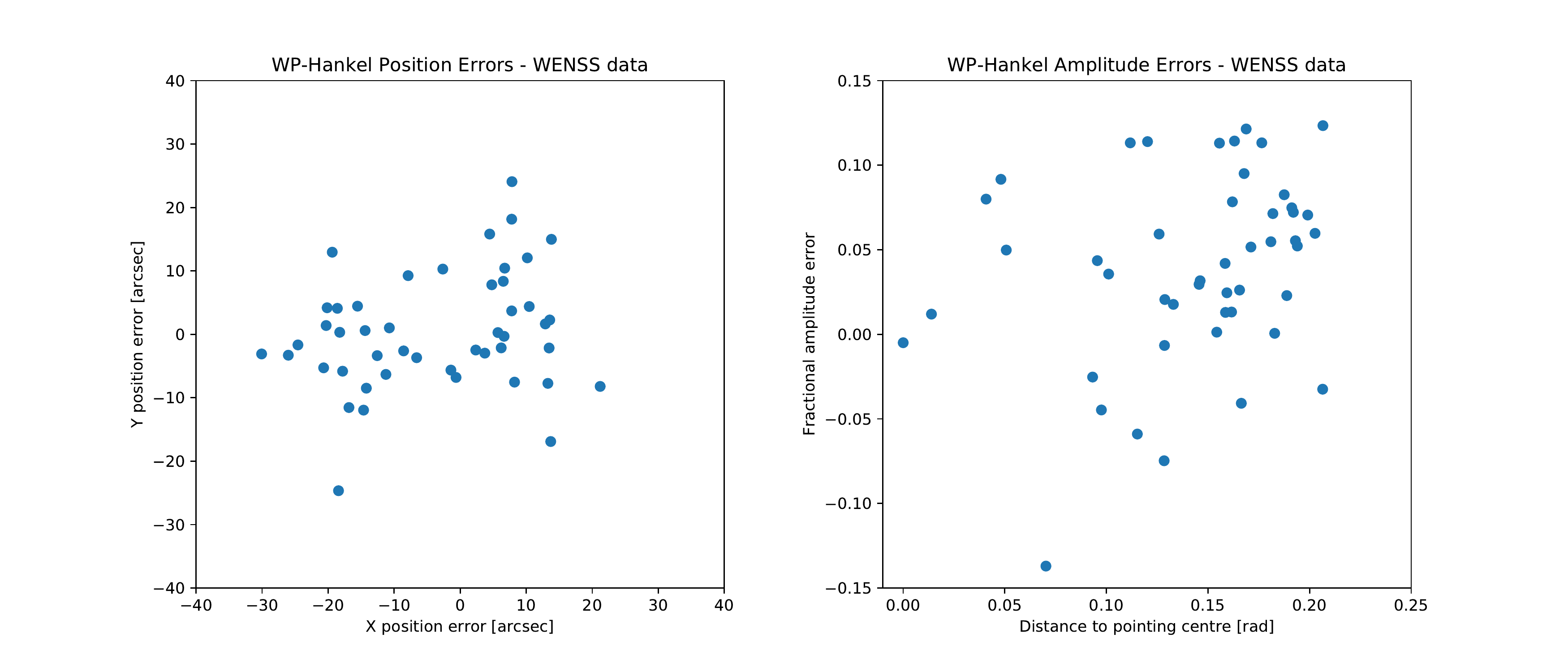}
\caption{Source position errors [arcsec] (left) and source peak amplitude errors [Jansky] (right) using the \emph{WP-Hankel} configuration, the 5DIAG measurement set and IM-S imager configuration (image size $2048 \times 2048$).}
\label{fig:wphankel_errors_wenss}
\end{figure*}

\begin{figure*}
\centering
\includegraphics[width=\textwidth]{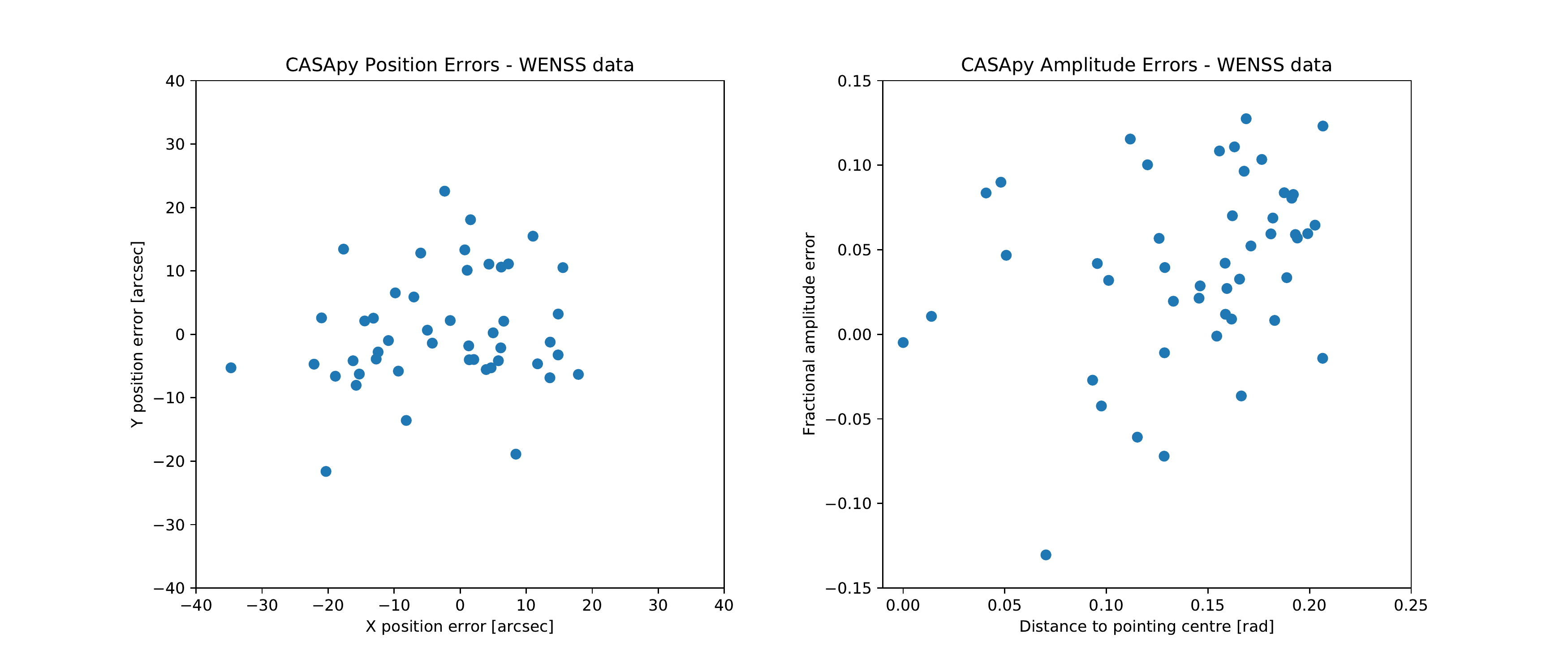}
\caption{Source position errors [arcsec] (left) and source peak amplitude errors [Jansky] (right) using the CASApy software with the 5DIAG measurement set and image size $2048 \times 2048$.}
\label{fig:casapy_errors_wenss}
\end{figure*}

In this section we compare the performance of the proposed $w$-projection algorithm with the widely used CASApy software\footnote{https://casa.nrao.edu} \citep{CASA}.
The results include the peak amplitude errors and position errors of each source detected in the output dirty image, using the STP source finder \citep{Lucas_SourceFind2019}.
We note that timing benchmarks are not performed here because CASApy has a high overhead in setting up the imager and writing the output results, which would make direct comparison unfair.

Regarding the algorithm settings, we used the default \emph{WP-FFT} and \emph{WP-Hankel} configuration as well as the IM-S imager configuration (image size $2048 \times 2048$). Regarding CASApy software, we tried to match its configuration with the one of the STP algorithm, from the few available parameters. We used the following settings:
\begin{lstlisting}
    gridmode=`widefield'
    niter = 0
    imsize = 2048
    cell=`60arcsec'
    wprojplanes = 128
\end{lstlisting}
where $niter=0$ is used to disable deconvolution.

Figure \ref{fig:wpfft_errors_5diag} presents the peak amplitude errors [Jansky] and position errors [arcsec] for each of the 5 sources of 5DIAG measurement set using the \emph{WP-FFT} configuration, \emph{WP-Hankel} configuration and the CASApy software, respectively.
In these results, we may observe that STP experiments seem to provide smaller errors in the position of the detected sources than CASApy. However, regarding the amplitude errors, there are slight differences which can favour STP or CASApy depending on the source. As expected, the peak amplitude error increases with the distance to the field centre, reaching up to an error of approximately 8\% for the farthest source (index 5 in the plot).
It is important to note that these results can be significantly improved by increased the number of $w$-planes used for $w$-projection in all algorithms. The main objective of these tests is to show that the developed algorithms perform with an accuracy performance comparable with a known imager implementation.

We also present the peak amplitude errors and position errors for the WENSS test data. These results are shown in Figures~\ref{fig:wpfft_errors_wenss}, \ref{fig:wphankel_errors_wenss} and \ref{fig:casapy_errors_wenss} for the \emph{WP-FFT} configuration, \emph{WP-Hankel} configuration and the CASApy software, respectively. Due to the fact that some sources are partially overlapped, not being distinguished by the source detection algorithm, we present the results only for 47 well detected sources in all 3 experiments. We can observe that position and amplitude errors present a similar distribution in all the figures, but some slight differences are visible between the CASApy and the STP plots. 
The average fractional amplitude error measured for \emph{WP-FFT}, \emph{WP-Hankel} and CASApy is 5.66 \%, 5.68 \% and 5.65 \%, respectively. These results demonstrate that the methods operate quite similarly, although some slight differences may exist due to different implementation details, numeric errors or different settings (\emph{e.g.} CASApy may not be using a Gaussian GCF).

\section{Conclusions}
\label{sec:conclusions}

Upcoming radio telescopes for wide field observations demand highly efficient imaging algorithms. 
In this paper we presented an optimised $w$-projection algorithm, which has been implemented and evaluated in the context of the SKA SDP STP software. The major difference between the proposed algorithm and the standard approach to $w$-projection is the use of a Hankel transform to speed-up the generation of the convolution kernels for wide-field imaging algorithms.

The experimental results presented here demonstrate the superior computational complexity of our proposed approach, which can generate convolution kernels 10 times faster than the standard 2D-FFT based approach. When analysing the impact of the proposed optimisation on global imager performance, we show that for a typical wide field imaging problem with image size of $16384\times 16384$ pixels, a speedup of $\sim 2$ is seen relative to standard $w$-projection on a 4-core machine. Furthermore, we demonstrate that this difference can be significantly larger when using an increased number of $w$-planes for improved output accuracy. This is due to the slow scaling of our proposed method with the number of $w$-planes, thus providing a computationally cheaper solution for higher output accuracy.

\section*{Acknowledgements}

The authors gratefully acknowledge support from the UK Science \& Technology Facilities Council (STFC).

%%%%%%%%%%%%%%%%%%%%%%%%%%%%%%%%%%%%%%%%%%%%%%%%%%

%%%%%%%%%%%%%%%%%%%% REFERENCES %%%%%%%%%%%%%%%%%%

% The best way to enter references is to use BibTeX:

\bibliographystyle{mnras}
\bibliography{fastw} % if your bibtex file is called example.bib

% Alternatively you could enter them by hand, like this:
% This method is tedious and prone to error if you have lots of references
%\begin{thebibliography}{99}
%\bibitem[\protect\citeauthoryear{Author}{2012}]{Author2012}
%Author A.~N., 2013, Journal of Improbable Astronomy, 1, 1
%\bibitem[\protect\citeauthoryear{Others}{2013}]{Others2013}
%Others S., 2012, Journal of Interesting Stuff, 17, 198
%\end{thebibliography}

%%%%%%%%%%%%%%%%%%%%%%%%%%%%%%%%%%%%%%%%%%%%%%%%%%

%%%%%%%%%%%%%%%%% APPENDICES %%%%%%%%%%%%%%%%%%%%%

%\appendix

%\section{Some extra material}

%If you want to present additional material which would interrupt the flow of the main paper,
%it can be placed in an Appendix which appears after the list of references.

%%%%%%%%%%%%%%%%%%%%%%%%%%%%%%%%%%%%%%%%%%%%%%%%%%

% Don't change these lines
\bsp	% typesetting comment
\label{lastpage}
\end{document}